# Un-renormalized Classical Electromagnetism

## M Ibison

Institute for Advanced Studies at Austin,

11855 Research Boulevard, Austin TX 78759-2443, USA

E-mail: ibison@earthtech.org

## Abstract

This paper follows in the tradition of direct-action versions of electromagnetism having the aim of avoiding a balance of infinities wherein a mechanical mass offsets an infinite electromagnetic mass so as to arrive at a finite observed value. Given that, in this respect the direct-action approached ultimately failed because its initial exclusion of self-action was found to be untenable in the relativistic domain, this paper continues the tradition considering instead a version of electromagnetism wherein mechanical action is excluded and self-action is retained. It is shown that the resulting theory is effectively interacting due to the presence of infinite forces. A vehicle for the investigation is a pair of classical point charges in a positronium-like arrangement for which the orbits are found to be self-sustaining and naturally quantized.

# 1. Introduction

In using the classical Maxwell theory to perform practical calculations one rarely deals explicitly with the infinite electromagnetic mass resulting from electromagnetic self-action. Implicitly, in the classical analogue of QED mass-renormalization, one assumes a negative infinite mechanical mass canceling the positive infinite electromagnetic mass of a point charge so as to arrive at the finite observed value. Though this balance of infinities may be unattractive, a success of the theory is a reasonable expression for the von Laue 4-vector, which can be obtained from the retarded action of a charged sphere upon itself - eventually letting the radius go to zero (see Boyer [1] and also Erber [2] for a review). The self-action that gives rise to the infinite electromagnetic mass therefore appears to be necessary to explain both radiation damping and reaction to acceleration.

Competing with the Maxwell theory is the direct-action version of classical electrodynamics (direct-action CED). The theory looked promising because it permitted the explicit exclusion of self-action, avoiding at the outset the infinite electromagnetic contribution to the mass [3-5]. One was then free to posit a finite mechanical mass without requiring a balance of infinities. The other distinguishing characteristic of the direct-action theory is that there are no vacuum degrees of freedom. As a result of the latter one then has the problem of somehow explaining the (observed) retarded field of a point source (having propagator $G_{ret} = G_+ + G_-$) by the collective action of multiple sources whose influence the theory demands propagate as $G_+$. Without that explanation there can be neither radiation nor von Laue 4-vector in the Newton-Lorentz equation of motion.

Wheeler and Feynman [6,7] made progress in favor of the direct-action theory by providing an explanation for the von Laue 4-vector as arising from the effects of distant absorbers on the forward light cone. In their version of direct-action EM the matter was treated classically, self-action was excluded, and future absorption explained retarded radiation and radiation reaction. The success of their explanation, however, depended on Cosmologies which are not currently favored [8-13]. Even so, the removal of infinities would seem to be a point in favor of direct-action CED over the Maxwell theory. Subsequently Hoyle and Narlikar [14] gave a quantum-mechanical version of the Wheeler-Feynman theory wherein the matter was treated by the Feynman path integral method. (Traditional quantization via a Hamiltonian is problematical because the action is in two times.) Davies [15] achieved the same goal using an S-matrix approach. In addition to demonstrating the emergence of retarded radiation, these works also gave, for example, the correct level shifts normally associated with the ZPF, i.e., processes traditionally regarded as mandating field quantization. (Pegg [16] gives a short calculation showing how the direct-action fields mimic the presence of the ZPF.)

Subsequently Feynman decided that the exclusion of self-action from CED was questionable because self-action was necessary for a satisfactory explanation of intermediate processes in QED [17]. He predicted that a consistent *relativistic* theory of direct action would, after all, have to retain self-action. That conclusion was supported in the comprehensive review by Pegg [10], and also by the author [11,18]. Feynman's suggestion was implemented by Hoyle and Narlikar in an extension of their earlier path integral work to the relativistic domain [19,20]. Similarly, Davies incorporated self-action in an extension to the relativistic domain of his earlier scattering matrix approach to matter quantization [21]. These works produced results in accord with QED at all perturbation orders, confirming the prediction by Feynman. But along with the return of the self-action came the usual problems requiring the usual techniques for dealing with divergent self-energy. In the end the direct-action theory necessitated the same delicate balance between infinities demanded of the Maxwell theory in order to arrive at the observed mass.

The undertaking reported here is a continuation of the search for an alternative to the balance of infinite masses. Its focus is a classical theory of EM wherein the action is entirely electromagnetic - without the usual term $I_{mech} = -m \int \sqrt{dx^2}$ for mechanical mass-action. It is clear from the foregoing that self-action must be retained in any new approach, independent of its Maxwell or direct-action heritage. The immediate consequence of such a modification therefore is that the total mass must be pure electromagnetic and infinite - uncompensated by mechanical mass.

The obvious advantage of 'uncompensated EM' is that it does not require a balance of infinities. Less apparent is how such a theory could overcome the obvious objection that charged particles with uncompensated electromagnetic mass apparently cannot ever effectively interact. But an EM theory with uncompensated mass turns out to provide for the existence of the infinite forces necessary to accelerate these infinite (electromagnetic) masses. The possibility of such forces may be inferred from the traditional expression for the Lorentz force tensor wherein, (using the notation of Jackson [22]), one observes the persistent motif of a denominator of the form $(1 - \mathbf{v}.\mathbf{n})$ raised to some positive power. Here **v** is the velocity of the source with respect to the reference frame, **n** is the unit vector from the source (traditionally at the retarded time) to the field point. The implication, if those expressions remain applicable, is that charges feel an infinite force in the event that they reside simultaneously on the light cone and the Cerenkov cone of another charge. (Here the word cone is being used in two different ways. The light cone is of co-dimension 1 in space-time, with symmetry axis parallel with the time direction. In 2+1 therefore, the light-cone is an ordinary 2-surface double-cone. The Cerenkov cone is also of co-dimension 1 in space-time, but has symmetry its axis parallel with **v** in the space 'plane' *t*=constant. The Cerenkov-cone is therefore an ordinary 2-surface double-cone in 3+0, but in 2+1 is a pair of planes.) A second objection may now be raised that, at the least, this requires that the latter move superluminally, which possibility is normally excluded by the traditional form of the action. However, classically, the restriction to subluminal motion is a consequence of the form of the mechanical mass action, and therefore that restriction is absent from uncompensated classical EM.

Given the re-habilitation of self-action in direct-action EM, both the Maxwell and direct action formulations are possible vehicles for the `technique' of non-compensation. In this paper we consider only the direct action version of the theory, mainly on the grounds of simplicity; outside of the particle interactions there are no incoming or outgoing radiation fields complicating the analysis. We investigate here a relativistic classical implementation of uncompensated EM, and leave aside for the present the possibility of an implementation of a relativistic quantum version along the lines of, for example Hoyle and Narlikar [19].

The remainder of the document is structured as follows. The next section presents the uncompensated version of direct-action CED. Self-action is obtained through a limiting procedure analogous to letting the size $\Delta$ of the point particle tend to zero from some initially finite value. Mathematically, it resembles the regularization method of the Feynman propagator $G_F$. Some care is required in deriving the Euler equation from the action; traditional derivations do not generally account for the possibility of superluminal motion, which becomes a possibility when there is no mechanical mass. A detailed derivation of the Euler equation is given in the appendix. The plausibility of the limiting procedure – of letting the size $\Delta$ tend to zero – is demonstrated by solving for the trivial case of uniform motion. In sections 3 and 4 is demonstrated the main claim of the paper, namely that the theory is effectively interacting despite the infinite self-energies. The vehicle for the demonstration is a pair of oppositely charged particles moving at constant speed on a self-sustaining circular path about a common origin. This positronium-like arrangement is then to be shown to be solution of the Euler equation. In section 3 is identified a principal constraint - the `Cerenkov cone condition' - in order that the arrangement be a solution. Further parameters of the motion are determined in section 4 to sufficient degree in $\Delta$ that the motion satisfies all the singular components of the Euler equation as $\Delta \to 0_+$. That is, the parameters are chosen so that the motion is at least consistent with the action of the singular forces. The total energy and angular momentum of the arrangement is computed in section 5, referring to the Appendix to provide general expressions for these quantities for a two-time action of which direct-action EM is a particular case. In section 6 there is some brief discussion of the results.

## 2. Uncompensated direct action EM

### 2.1 Notation

To reduce reliance on indexes and improve readability 4-vectors (which here are not necessarily Lorentz vectors) are denoted by a lower-case Latin symbol, $x \equiv \{x^a\} = (x_0, x_1, x_2, x_3)$, where the spatial basis is Cartesian, and where the default, when the vector appears alone, is the Lorentz contravariant form. The scalar product of two vectors $x$ and $y$ is written $x \circ y \equiv x^a y_a$ (vectors appearing to the right of the operator $\circ$ are in Lorentz covariant form). The anti-symmetric product is denoted by $x \wedge y \equiv \{x^a y^b - y^a x^b\}$. The second rank force tensor is denoted by $F \equiv \{F^{ab}\}$, with the right scalar product $F \circ u$ taken to mean $\{F^{ab} u_b\}$. $\lambda$ and $\lambda'$ denote a pair of ordinal parameters for the trajectories of point particles; $x_j \equiv x_j(\lambda)$ represents the 4-position of particle $j$ at ordinal time $\lambda$, and $x'_k \equiv x_k(\lambda')$ represents the 4-position of particle $k$ at ordinal time $\lambda'$. $s_{k,j} \equiv x_k(\lambda') - x_j(\lambda)$ is the 4-vector difference at the two times. $u_j \equiv dx_j/d\lambda$ and $a_j = d^2 x_j/d\lambda^2$ is the `ordinal 4-velocity' and `ordinal 4-acceleration' of the trajectory $j$ at time $\lambda$, and similarly for the primed quantities.

### 2.2 Action

Using the notation introduced above the traditional formulation of classical direct-action EM for a collection of charged particles with rest mass $m$ is

$$I = -\frac{1}{2} \sum_{\substack{j,k \\ j \neq k}} e_j e_k \int d\lambda \int d\lambda' u_j \circ u'_k \delta(s_{k,j}^2) - \sum_j m_j \int d\lambda \sqrt{u_j^2} \qquad (1)$$

(see for example [7]). Self-action is excluded by excluding the diagonal terms $j = k$ from the double sum. By contrast, in the uncompensated direct-action EM framework under consideration here, we admit electromagnetic self-action and deny additional mechanical action. That is, we consider the electrodynamics of completely uncompensated point charges formally possessing infinite electromagnetic mass and zero mechanical mass. Correspondingly, (1) is initially replaced with simply

$$I = -\sum_{j,k} e_j e_k \int d\lambda \int d\lambda' u_j \circ u'_k \delta(s_{k,j}^2). \qquad (2)$$

(The factor of ½ is no longer of any consequence and is omitted for convenience.) The contribution to the action at the point $\lambda = \lambda'$ when $j = k$ will be referred to as the *local* self-action. Because this contribution is singular, we immediately replace (2) with

$$I(\Delta^2) = -\sum_{\sigma_s=\pm 1}\sum_{j,k} e_j e_k \int d\lambda \int d\lambda' u_j \circ u'_k \delta(s^2_{k,j} + \sigma_s \Delta^2) = -2\Delta^2 \sum_{j,k} e_j e_k \int d\lambda \int d\lambda' u_j \circ u'_k \delta(s^4_{k,j} - \Delta^4) \quad (3)$$

and assume that the physics of uncompensated direct action EM is formally described by the Euler equation for (3) in the limit that $\Delta \to 0_+$. (Always it will be assumed that $\Delta$ is positive.) That is, (3) is first extremized without regard to the magnitude of $\Delta$, and only subsequently does one let $\Delta \to 0_+$. This procedure retains the Lorentz invariance and parameterization-invariance of (1) and (2). However, unlike (2), though like (1), the procedure is not scale invariant. There exist other plausible limiting procedures.

The form of (1) permits only sub-luminal speeds lest the mechanical action becomes imaginary. Since this restriction is absent in (3) both superluminal speeds and time-reversals are allowed. It follows that in (3) the assumption that $t(\lambda) \equiv x_0(\lambda)$ is a monotonic function amounts to the imposition of an external constraint on the motion. In general, in using (3) one cannot impose a relationship between the coordinate time $x_0(\lambda)$ and the ordinal time $\lambda$ without prior physical justification. (It is worth recalling that a superluminal trajectory without time-reversals with respect to some reference frame $\Sigma$, say, will acquire time reversals when viewed from some inertial frames that are moving *sub*-luminally with respect to $\Sigma$.)

As a result of the parameterization invariance the Euler equation is form invariant under a Lorentz transformation even if a parameterization is chosen such that $\lambda$ is not a Lorentz scalar (e.g. $\lambda$ is not the proper-time) and even though, in that case, *u* and *a* will not be Lorentz vectors and scalar products such as $u \circ a$ and $u^2 \equiv u \circ u$ will not be Lorentz scalars. In this paper will be considered only candidate solutions to the Euler equation that do not time-reverse when viewed from the laboratory frame, and it will be convenient to exploit the freedom of parameterization to choose $\lambda = t$.

The light cone condition $s^2_{k,j} = 0$ in (1) and (2) that picks out the times of electromagnetic interaction is replaced in (3) by the `modified light-cone' condition $s^2_{k,j} \pm \Delta^2 = 0$. Actually the double cone of co-dimension 1 in 3+1D is replaced with three disjoint surfaces each of co-dimension 1. In general this will give rise to interactions as illustrated in the 1+1 D cross section, Figure 1a. The interval *s* is time-like if $j = k$ and the particle in question is always subluminal. Otherwise *s* is can be time-like, space-like or null, in which case both signs of $\sigma_s$ need be retained. Note however that one or more of the interactions will disappear in the particular case that the distant interacting trajectory strikes the light-cone of the local particle at a tangent. This possibility is illustrated in Figure 1b for motion in 1+1D, and will turn out to be important in the subsequent analysis of superluminal dual circular motion in 2+1D.

## 2.3 Euler equation

Since there is no mechanical mass, variation of (3) gives simply that the particle moves to ensure that the total force is zero,

$$f_j(\lambda) = \sum_k e_j F_{k,j}(\lambda) \circ u_j(\lambda) = 0. \tag{4}$$

$f_j(\lambda)$ is a force 4-vector and $F_{k,j}(\lambda)$ is the force tensor derived from the trajectories of all the particles, including the particle $j = k$. The derivation of $F$ from the Euler equation involves a few departures from the traditional derivation from the Liénard-Wiechert potential, for example as given by Jackson [22]. The Euler equation for a general two-time action is derived in the appendix. Inserting into the result (A16) the particular expression for the direct-action kernel (A6), one obtains, $\forall l$,

$$\sum_{j,k}\sum_{\sigma_s=\pm 1} e_j e_k \int d\lambda' \left( \frac{\partial}{\partial x_l} - \frac{d}{d\lambda}\frac{\partial}{\partial u_l} \right) u_j \circ u'_k \delta\left(s_{k,j}^2 + \sigma_s \Delta^2\right) = 0. \tag{5}$$

It is to be understood that in the above we intend to take the limit $\Delta \to 0$ after performing the integration. Carrying out the differentiations in (5) one has

$$\sum_k \sum_{\sigma_s=\pm 1} e_j e_k \int d\lambda' \, 2\left(u'_k(u_j \circ s_{k,j}) - (u_j \circ u'_k)s_{k,j}\right) \delta'\left(s_{k,j}^2 + \sigma_s \Delta^2\right) = 0, \tag{6}$$

and writing

$$\delta'\left(s_{k,j}^2 + \sigma_s \Delta^2\right) = \frac{1}{2(u'_k \circ s_{k,j})} \frac{d}{d\lambda'} \delta\left(s_{k,j}^2 + \sigma_s \Delta^2\right) \tag{7}$$

the Euler equation becomes

$$\begin{aligned}&\sum_k \sum_{\sigma_s=\pm 1} e_j e_k \int d\lambda' \, \frac{u'_k(u_j \circ s_{k,j}) - (u_j \circ u'_k)s_{k,j}}{u'_k \circ s_{k,j}} \frac{d}{d\lambda'}\delta\left(s_{k,j}^2 + \sigma_s \Delta^2\right) = 0 \\ &\Rightarrow \sum_k \sum_{\sigma_s=\pm 1} e_j e_k \int d\lambda' \, \frac{d}{d\lambda'}\left(\frac{u'_k \wedge s_{k,j}}{u'_k \circ s_{k,j}}\right) \circ u_j \delta\left(s_{k,j}^2 + \sigma_s \Delta^2\right) = 0\end{aligned}. \tag{8}$$

Performing the integration over $\lambda'$, this is

$$\sum_k \sum_{\sigma_s=\pm 1} \sum_{\text{roots}} e_j e_k \frac{1}{|u'_k \circ s_{k,j}|} \frac{d}{d\lambda'}\left(\frac{u'_k \wedge s_{k,j}}{u'_k \circ s_{k,j}}\right) \circ u_j = 0 \tag{9}$$

where the roots are the $\lambda'(\lambda)$ such that $s_{k,j}^2 + \sigma_s \Delta^2 = 0$ (to be substituted into the kernel of the sum *after* the differentiation has been performed). Comparing with (4) one sees that the force tensor is

$$F_{k,j}(\lambda) = \sum_{\sigma_s=\pm 1} \sum_{\text{roots}} e_k \frac{1}{|u'_k \circ s_{k,j}|} \frac{d}{d\lambda'}\left(\frac{u'_k \wedge s_{k,j}}{u'_k \circ s_{k,j}}\right). \tag{10}$$

Eq. (10) gives the generalization of the traditional EM force tensor derived from the Liénard-Wiechert potential. Computing the derivatives, the 4-force on the $j^{th}$ particle is

$$f_j = \sum_k \sum_{\sigma_s = \pm 1} \sum_{\text{roots}} e_j e_k \frac{1}{|u'_k \circ s_{k,j}|^3} \begin{pmatrix} \left((u'_k \circ s_{k,j})(a'_k \circ u_j) - (u'^2_k + a'_k \circ s_{k,j})(u'_k \circ u_j)\right) s_{k,j} \\ + (u'^2_k + a'_k \circ s_{k,j})(u_j \circ s_{k,j}) u'_k - (u'_k \circ s_{k,j})(u_j \circ s_{k,j}) a'_k \end{pmatrix}. \tag{11}$$

## 2.4 Uniform motion

We first briefly consider a single particle executing uniform motion in space. Because there is only one particle the indexes $j$ and $k$ can be omitted. Since there are no coordinate time reversals we can exercise the parameterization invariance to choose the ordinary time: $x_0(\lambda) = \lambda$. Then $x = \lambda(1, \mathbf{v})$, where $\mathbf{v}$ is the ordinary 3-vector velocity, and $s_{k,j} = (\lambda' - \lambda)(1, \mathbf{v})$. All second derivatives appearing in (11) are zero, and therefore

$$f = e^2 \sum_{\sigma_s = \pm 1} \sum_{\lambda'(\lambda) \text{ st } s^2 + \sigma_s \Delta^2 = 0} \frac{u'^2}{|u' \circ s|^3} \left((u \circ s) u' - (u' \circ u) s\right). \tag{12}$$

The $\Delta$-modified light-cone condition is

$$s^2 + \sigma_s \Delta^2 = 0 \Rightarrow (\lambda' - \lambda)^2 (1 - \mathbf{v}^2) + \sigma_s \Delta^2 = 0. \tag{13}$$

In the case that $|\mathbf{v}| < 1$ there are roots corresponding to $\sigma_s = -1$ but not for $\sigma_s = +1$. The situation is reversed if $|\mathbf{v}| > 1$. Since the force is a sum over both contributions, one can set

$$|\lambda' - \lambda| = \Delta \big/ \sqrt{|1 - \mathbf{v}^2|}. \tag{14}$$

Using this, the ratio in the sum in (12) is

$$\frac{u'^2}{|u' \circ s|^3} = \frac{(1 - \mathbf{v}^2)^2}{|\lambda' - \lambda|^3 |1 - \mathbf{v}^2|^3} = \frac{\sqrt{|1 - \mathbf{v}^2|}}{\Delta^3}. \tag{15}$$

The term in parentheses in (12) is identically zero, and therefore, following the prescription of taking the limit $\Delta \to 0_+$ after forming the Euler equation - the self-force is unambiguously zero. It is concluded that all uniform motions are valid solutions of the Euler equation.

It may be observed that if the delta function $\delta(s^2_{k,j} - \Delta^2)$ in (3) were absent only super-luminal roots would exist. Formally this would mean that *any* sub-luminal motion of a single particle would solve the Euler equation (which is then already zero). Likewise, if the delta function $\delta(s^2_{k,j} + \Delta^2)$ were absent, *any* superluminal motion of a single particle confined to a single space-time plane (i.e. in one space dimension) would generate no roots, and therefore would also formally be a solution of the Euler equation. It follows that both delta functions together ensure that arbitrary motions are not automatic solutions of the Euler equation.

## 3. Superluminal circular motion of two charges

### 3.1 Specification of the motion

We now explore as a possible solution of the Euler equation (4) two charges in circular motion, both moving with superluminal speeds with respect to the laboratory reference frame. Since the Euler equation is parameterization invariant and there are no time reversals of the superluminal motion when referred to the (static) laboratory frame it is again permissible and convenient to parameterize the trajectories with the laboratory time, $\lambda = t$. Again, the four-component vectors $x$, $u$, $a$, etc. are not Lorentz vectors and their scalar products are not Lorentz invariants. Accordingly we let $j,k \in \{1,2\}$ and

$$u_j \equiv u_j(t) = (1, \mathbf{v}_j(t)), \quad a_j \equiv a_j(t) = (0, \mathbf{a}_j(t))$$
$$u'_k \equiv u_k(t') = (1, \mathbf{v}_k(t')), \quad a'_k \equiv a_k(t') = (0, \mathbf{a}_k(t'))$$
(16)

The modified light cone condition is

$$s_{k,j}^2(t',t) + \sigma_s \Delta^2 = 0 \Rightarrow t' = \{t'_{l,k,j}(t; \Delta, \sigma_s)\}$$
(17)

where (for each $j$, $k$) $l$ enumerates the roots $t'$. We restrict consideration to two particles in circular motion about a common origin. Based on symmetry grounds – that each particle much generate the force necessary to sustain the other in circular motion at all times – it is reasonable that the two particles must have the same speed and must be $\pi$ radians out of phase in the laboratory frame. This configuration turns out to be sufficient for stability, though other possibilities are not ruled out here. Confining the motion to the $x_1, x_2 = x, y$ plane and suppressing the $x_3 = z$ coordinate one has, for $j \in [1,2]$,

$$x_j(t) = (t, r\cos(\omega t + \phi_j), r\sin(\omega t + \phi_j))$$
(18)

where here and throughout $\omega$ and $r$ are fixed positive quantities, and $\phi_2 - \phi_1 = \pi$. With this the other 4-vectors are

$$s_{k,j}(t',t) = x_k(t') - x_j(t) = (t'-t, r\cos(\omega t' + \phi_k) - r\cos(\omega t + \phi_j), r\sin(\omega t' + \phi_k) - r\sin(\omega t + \phi_j))$$
$$u_j(t) = (1, -|\mathbf{v}|\sin(\omega t + \phi_j), |\mathbf{v}|\cos(\omega t + \phi_j))$$
$$a_j(t) = -|\mathbf{v}|\omega(0, \cos(\omega t + \phi_j), \sin(\omega t + \phi_j))$$
(19)

with $u'_k$ and $a'_k$ obtained simply by argument substitution. The modified light cone condition (17) is then

$$(t'_{l,k,j} - t)^2 + 2r^2 \left(\cos\left(\omega(t'_{l,k,j} - t) + \phi_k - \phi_j\right) - 1\right) + \sigma_s \Delta^2 = 0.$$
(20)

Observing that the phase difference is 0 or π, define

$$\sigma_{k,j} = 2\delta_{k,j} - 1 = \begin{cases} 1 & \text{if } k = j \\ -1 & \text{if } k \neq j \end{cases} \tag{21}$$

and also define the dimensionless quantities

$$\theta_{l,k,j} = \omega(t'_{l,k,j} - t), \quad \varepsilon = \omega\Delta/2 \tag{22}$$

where $\varepsilon$ is positive because $\omega$ and $\Delta$ are positive. With (21) and (22), (20) can be written

$$\theta_{l,k,j}^2 + 2\mathbf{v}^2\left(\sigma_{k,j}\cos\theta_{l,k,j} - 1\right) + 4\sigma_s\varepsilon^2 = 0. \tag{23}$$

Because (23) is even in $\theta$, for fixed $j,k$ the $\theta_{l,k,j}$ come in opposite signed pairs, making it convenient to employ a signed index $l$ such that

$$\theta_{l,k,j} \geq 0, \quad \theta_{-l,k,j} = -\theta_{l,k,j}. \tag{24}$$

This symmetry is the expected outcome of the configuration of the two charges: for every intersection on the *future* (retarded) light cone at relative phase $\theta$ there exists another intersection on the *past* (advanced) light cone from the same point at the relative phase $-\theta$.

### 3.2 Forces on the particles

Since the light cone condition depends only on $\theta$ it is convenient to go to the rotating frame of the $j^{th}$ particle in whose frame the force is to be computed:

$$x_\mu \to x_{\bar{\mu}} = \begin{pmatrix} 1 & 0 & 0 \\ 0 & \cos(\omega t + \phi_j) & \sin(\omega t + \phi_j) \\ 0 & -\sin(\omega t + \phi_j) & \cos(\omega t + \phi_j) \end{pmatrix} x_\mu. \tag{25}$$

In this frame the quantities (18) and (19) are

$$\begin{aligned} s_{k,j} &\to \frac{1}{\omega}\left(\theta, |\mathbf{v}|(\sigma_{k,j}\cos\theta - 1), \sigma_{k,j}|\mathbf{v}|\sin\theta\right) \\ u'_k &\to \left(1, -\sigma_{k,j}|\mathbf{v}|\sin\theta, \sigma_{k,j}|\mathbf{v}|\cos\theta\right) \\ a'_k &\to -\sigma_{k,j}\omega|\mathbf{v}|(0, \cos\theta, \sin\theta) \\ u_j &\to (1, 0, |\mathbf{v}|) \end{aligned} \tag{26}$$

The scalar products appearing in the force, Eq. (11), are frame independent, and can be computed using either (26) or (18) and (19). Suppressing the indexes on $\theta$ one obtains

$$\begin{aligned} u'^2_k &= 1 - \mathbf{v}^2 \\ a'_k \circ s_{k,j} &= \mathbf{v}^2\left(1 - \sigma_{k,j}\cos\theta\right) \\ u'_k \circ u_j &= u'^2_k + a'_k \circ s_{k,j} = 1 - \sigma_{k,j}\mathbf{v}^2\cos\theta \\ u_j \circ s_{k,j} &= u'_k \circ s_{k,j} = \frac{1}{\omega}\left(\theta - \sigma_{k,j}\mathbf{v}^2\sin\theta\right) \\ a'_k \circ u_j &= \sigma_{k,j}\mathbf{v}^2\omega\sin\theta \end{aligned} \tag{27}$$

Noting the equalities in (27) of some of the terms in (11), the force in the rotating frame is

$$f_j \to \sum_{\sigma_s=\pm 1} \sum_{l,k} e_k e_j \frac{\left((u'_k \circ s_{k,j})(a'_k \circ u_j) - (u'_k \circ u_j)^2\right) s_{k,j} + (u'_k \circ s_{k,j})\left((u'_k \circ u_j) u'_k - (u'_k \circ s_{k,j}) a'_k\right)}{|u'_k \circ s_{k,j}|^3}, \qquad (28)$$

where the sum is over the roots $l$ of $\theta$ as given by (23) and (24). As a consequence of the latter only terms that are even in $\theta$ survive the sum. Using (27) let us denote the parity of the scalar products in (28) by the subscripts $o$ and $e$, corresponding respectively to odd and even:

$$f_j = \sum_{\sigma_s=\pm 1} \sum_{l,k} e_k e_j \frac{\left((u'_k \circ s_{k,j})_o (a'_k \circ u_j)_o - (u'_k \circ u_j)^2_e\right) s_{k,j} + (u'_k \circ s_{k,j})_o \left((u'_k \circ u_j)_e u'_k - (u'_k \circ s_{k,j})_o a'_k\right)}{|u'_k \circ s_{k,j}|^3_e}. \qquad (29)$$

Letting $\hat{O}$ and $\hat{E}$ extract the parts of their operands that are respectively odd and even in $\theta$ it is deduced that (29) can be written

$$f_j = \sum_{\sigma_s=\pm 1} \sum_{l>0} \sum_{k} 2 e_k e_j \frac{1}{|u'_k \circ s_{k,j}|^3_e}$$
$$\times \left(\left((u'_k \circ s_{k,j})_o (a'_k \circ u_j)_o - (u'_k \circ u_j)^2_e\right) \hat{E}(s_{k,j}) + (u'_k \circ s_{k,j})_o \left((u'_k \circ u_j)_e \hat{O}(u'_k) - (u'_k \circ s_{k,j})_o \hat{E}(a'_k)\right)\right) \qquad (30)$$

With reference to (27) it is seen that the 0$^{th}$ and 2$^{nd}$ (the time and $x_2$- direction) components of the force are automatically zero, which is a predictable consequence of the symmetry of the configuration. This leaves only $x_1$ component of the force to be resolved. Noting that

$$(u'_k \circ s_{k,j})(a'_k \circ u_j) - (u'_k \circ u_j)^2 = (\theta - \sigma_{k,j} \mathbf{v}^2 \sin\theta)\sigma_{k,j} \mathbf{v}^2 \sin\theta - (1 - \sigma_{k,j} \mathbf{v}^2 \cos\theta)^2$$
$$= \sigma_{k,j} \mathbf{v}^2 (2\cos\theta + \theta \sin\theta) - \mathbf{v}^4 - 1 \qquad (31)$$

and letting $\hat{\mathbf{x}}$ be a unit vector in the $x_1$- direction, the total force on the $j^{th}$ particle is

$$\mathbf{f}_j = \sum_{\sigma_s=\pm 1} \sum_{l>0} \sum_{k} \frac{2 e_k e_j \omega^2 \hat{\mathbf{x}}}{|\theta - \sigma_{k,j} \mathbf{v}^2 \sin\theta|^3} \begin{pmatrix} \left(\sigma_{k,j} \mathbf{v}^2 (2\cos\theta + \theta\sin\theta) - \mathbf{v}^4 - 1\right)\left(\sigma_{k,j}|\mathbf{v}|\cos\theta - |\mathbf{v}|\right) \\ + (\theta - \sigma_{k,j}\mathbf{v}^2 \sin\theta)\begin{pmatrix} -(1 - \sigma_{k,j}\mathbf{v}^2 \cos\theta)\sigma_{k,j}|\mathbf{v}|\sin\theta \\ +(\theta - \sigma_{k,j}\mathbf{v}^2 \sin\theta)\sigma_{k,j}|\mathbf{v}|\cos\theta \end{pmatrix} \end{pmatrix}$$
$$= \sum_{\sigma_s=\pm 1} \sum_{l>0} \sum_{k} \frac{2 e_k e_j \omega^2 |\mathbf{v}| \hat{\mathbf{x}}}{|\theta - \sigma_{k,j}\mathbf{v}^2 \sin\theta|^3} \begin{pmatrix} \mathbf{v}^4 + \mathbf{v}^2 + 1 + \mathbf{v}^2 \cos^2\theta + (\theta^2 - \mathbf{v}^4 - 2\mathbf{v}^2 - 1)\sigma_{k,j}\cos\theta \\ -(\mathbf{v}^2 + 1)\sigma_{k,j}\theta\sin\theta \end{pmatrix} \qquad (32)$$
$$= \sum_{\sigma_s=\pm 1} \sum_{l>0} \sum_{k} \frac{2 e_k e_j \omega^2 |\mathbf{v}| \hat{\mathbf{x}}}{|\theta - \sigma_{k,j}\mathbf{v}^2 \sin\theta|^3} \begin{pmatrix} (\mathbf{v}^2 + 1)^2 + \left(\theta^2 - (\mathbf{v}^2 + 1)^2\right)\sigma_{k,j}\cos\theta \\ -(\mathbf{v}^2 + 1)\sigma_{k,j}\theta\sin\theta - \mathbf{v}^2 \sin^2\theta \end{pmatrix}$$

(the time- component of the force – the power – is already zero). Restoring the indexes on $\theta$ one can write $\mathbf{f}_j = \tilde{f}_j \mathbf{f}_{norm}$ where $\tilde{f}_j$ is the scalar

$$\tilde{f}_j = \sum_{\sigma_s=\pm 1}\sum_{l>0}\sum_{k}\mathrm{sgn}(e_1 e_2)\frac{\begin{pmatrix}\left(\mathbf{v}^2+1\right)^2+\left(\theta_{l,k,j}^2-\left(\mathbf{v}^2+1\right)^2\right)\sigma_{k,j}\cos\theta_{l,k,j}\\-\left(\mathbf{v}^2+1\right)\sigma_{k,j}\theta_{l,k,j}\sin\theta_{l,k,j}-\mathbf{v}^2\sin^2\theta_{l,k,j}\end{pmatrix}}{\left|\theta_{l,k,j}-\sigma_{k,j}\mathbf{v}^2\sin\theta_{l,k,j}\right|^3} \qquad (33)$$

and where

$$\mathbf{f}_{norm} \equiv 2e^2\omega^2\left|\mathbf{v}\right|\hat{\mathbf{x}} \qquad (34)$$

is a fixed finite force independent of $j$, $k$ and $l$.

### 3.3 Classification of forces

A superluminally-moving particle may interact with itself by crossing its own light cone any number of times. The first such crossing is `local' in that the associated time interval $t'-t \to 0$ as $\varepsilon \to 0$. Other light cone crossings wherein $t'-t \neq 0$ at $\varepsilon = 0$ will be referred to as `distant'. Whilst the local crossing necessarily connotes self-interaction, distant crossings may be either of `one's own' light cone, or of the light cone of another particle. In addition to these distinctions it is also important to distinguish between forces that are singular from those that are finite as $\varepsilon \to 0_+$. It will be seen below that the local self-force is necessarily singular, whereas the distant self-force and distant force from other particles can be either finite or singular. It will turn out that there are different degrees of singularity, though it will be sufficient for the following discussion to distinguish simply between finite and singular forces. In summary then, a light-cone crossing connoting an electromagnetic interaction and giving rise to a force can be characterized by three qualities, each of which can take two values: $strength \in \{finite, singular\}$, $proximity \in \{local, distant\}$, and $source \in \{self, not\text{-}self\}$.

We have already identified an intrinsic force that, in terms of this categorization scheme, has the quality vector $(singular, local, self)$. In accord with the labels introduced in section 3.1, for particle 1 the $source = self$ quality gives that $j = k = 1$. If we label the roots of (23) with $l$ increasing as one moves from *local* to *distant*, then the force in question will be associated with $l = 1$, and therefore with the angle $\theta_{1,1,1}$. Taking into account that $j = k$ and (24), and making the substitution $\theta_{1,1,1} = 2\alpha$, one has that $\alpha$ satisfies

$$\alpha^2 - \mathbf{v}^2\sin^2\alpha + \sigma_s\varepsilon^2 = 0, \quad \alpha > 0, \quad \lim_{\varepsilon\to 0_+}\alpha = 0. \qquad (35)$$

(Recall that the restriction to positive $\alpha$ is in accord with the derivation of (30) from (29).) In Figure 1b the two points of interaction corresponding to both signs of $\alpha$ are shown as two red points close to the origin, the latter being the 'present' position of the particle. Since the particle is superluminal (35) has real solutions only for positive $\sigma_s$, as is apparent from the figure. With this, Eq. (33) gives that the local force on particle 1 is:

$$\tilde{f}_{local} = \frac{\left(\mathbf{v}^2 + 1\right)^2 + \left(4\alpha^2 - \left(\mathbf{v}^2 + 1\right)^2\right)\cos 2\alpha - \left(\mathbf{v}^2 + 1\right)2\alpha \sin 2\alpha - \mathbf{v}^2 \sin^2 2\alpha}{\left|2\alpha - \mathbf{v}^2 \sin 2\alpha\right|^3} \tag{36}$$

where there are now no sums and $\alpha$ has just one value satisfying, for fixed speed,

$$\alpha^2 - \mathbf{v}^2 \sin^2 \alpha + \varepsilon^2 = 0, \quad \alpha > 0, \quad \lim_{\varepsilon \to 0_+} \alpha = 0. \tag{37}$$

The subscript *local* will turn out to be sufficient to distinguish the force in future calculations. In the following section it is shown that this force can be identified with (-1 times) the mass-acceleration of Newton's second law. Since it is *singular*, the associated mass is infinite, and one concludes that this force offers infinite resistance to the action of external forces.

In order to sustain circular orbits the charges must be subject to another force whose strength is *singular*. Traditionally a force having a quality vector other than (*singular*,*local*,*self*) is termed 'external'. However, that designation will not be used here because such a force might mistakenly be interpreted as having a source that is necessarily *not-self*. By contrast here there exists the possibility that other forces (i.e. with qualities other than (*singular*,*local*,*self*)) can be sourced either by *self* or by *not-self*, the former possibility arising because the charges are superluminal. The sought for additional *singular* force may even have multiple *singular*, *distant* contributions from either or both (*self* and *not-self*) particles. In this document however we will exclusively consider an arrangement wherein just one, other, singular force is sourced exclusively by the other, oppositely charged, particle. Such a force therefore has quality (*singular*,*distant*,*not - self*), which force, acting on particle 1, in accord with the labels introduced in section 3.1, has labels $j = 1$ and $k = 2$. The label $l$ for the associated angle is the particular unique index, $l = L$ say, into the ordered set of roots that are the solutions of (23) for which the corresponding force is singular. Taking into account that $j \neq k$ and (24), and making the substitution $\theta_{L,2,1} = 2\beta$, one has that $\beta$ satisfies

$$\beta^2 - \mathbf{v}^2 \cos^2 \beta + \sigma_s \varepsilon^2 = 0, \quad \beta > 0. \tag{38}$$

(Again, recall that the restriction to positive $\beta$ is in accord with the derivation of (30) from (29).) Figure 1a shows the general case for arbitrary motion of a distant trajectory. The four blue points are the points of interaction between two distant trajectory segments and the present location (nominally of particle *j*) at the origin. However, in section 3.5 it is shown that the distant trajectory must strike the light cone of the local particle at a tangent. Figure 1b is a representation of this situation but where the motion is confined to 1+1D. With this additional constraint it is deduced that (38) can have a real solution for just one value of $\sigma_s$ in the region of some intersection point. (There may be other intersection points involving the same trajectory.) Accordingly we can drop the sum over $\sigma_s$ from (33) so that the distant force on particle 1 from particle 2 is

$$\tilde{f}_{distant} = \text{sgn}(e_1 e_2) \frac{\left(\mathbf{v}^2+1\right)^2 - \left(4\beta^2 - \left(\mathbf{v}^2+1\right)^2\right)\cos 2\beta + \left(\mathbf{v}^2+1\right)2\beta \sin 2\beta - \mathbf{v}^2 \sin^2 2\beta}{\left|2\beta + \mathbf{v}^2 \sin 2\beta\right|^3}. \qquad (39)$$

The subscript *distant* will turn out to be sufficient to identify this force in future calculations, though a full description is $(singular, distant, not\text{-}self)$.

Under the stated assumptions these two cases are an exhaustive catalogue of the singular forces. One also has, however, additional - finite - electromagnetic forces that are traditionally admitted in the classical and quantum analyses after subtraction of the electromagnetic mass. For the system under discussion it will become apparent that, unlike the case of *singular* forces, there are a variable number of contributions to the total finite force from both particles, these coming from multiple distant crossings of the light cone (see Figure 3b). The variability considerably complicates an already lengthy analysis, and so this document will concern itself only with balance of the singular forces. This is not a big loss however, because (as will become clear) none of the quantities of interest – the angle $\beta$ (at $\varepsilon = 0$), the speed $|\mathbf{v}|$, energy, and angular momentum – depend, to leading order, on the balance of the finite forces.

Since there is no mechanical mass the Euler equation (4) dictates that the sum of the electromagnetic forces on each particle must vanish. Considering just the singular forces (36) and (39) one therefore has simply

$$\tilde{f}_{local} + \tilde{f}_{distant} = 0. \qquad (40)$$

Eqs. (35), (38) and (40) (with definitions (36), and (39)) are three equations in three unknowns $\alpha, \beta, \mathbf{v}^2$, to be solved as $\varepsilon \to 0_+$. In sections 3.3 and 3.4 we solve these only to the degree required to permit i) identification of the local contribution to the electromagnetic mass to order $1/\varepsilon$, and ii) identification of the condition that the distant force is singular to *some* degree. Determination of the latter condition is necessary but not sufficient to guarantee satisfaction of the Euler equation at the order $1/\varepsilon$. However, its determination is sufficient to identify an important geometrical constraint leading to quantization of the orbit. More precise determination of the dependencies of $\alpha, \beta, \mathbf{v}^2$ on $\varepsilon$ are necessary to prove satisfaction of the Euler equation up to but not including order $\varepsilon^0$. They are also necessary in order to compute the total energy and angular momentum of the system. The relevant calculations are given in section 4.

### 3.4 Local force to order $1/\varepsilon$

Eq. (35) permits the solution $\alpha \to 0_+$ as $\varepsilon \to 0_+$ which is the *local* solution that gives rise to local self-action responsible for self-force and self-energy. Note that $\alpha$ is proportional to $\varepsilon$ in approaching the limit. It is inferred from (35) and (38) that the speed is a function of $\varepsilon$. Writing $\mathbf{v}_0^2 \equiv \mathbf{v}^2\big|_{\varepsilon=0}$ (35) gives

$$\alpha = \frac{\varepsilon}{\sqrt{\mathbf{v}_0^2 - 1}} \tag{41}$$

plus higher-order terms in $\varepsilon$ whose presence here is suppressed. The associated self-force is found by inserting (41) into (36). Since $\alpha$ is small, the trig terms need be expanded only to lowest order in $\alpha$. The numerator in (36) is even in $\alpha$, but the 0$^{th}$ order – constant – term vanishes. One then has

$$\tilde{f}_{local} \to \frac{4\alpha^2 + 2\alpha^2(\mathbf{v}^2+1)^2 - 4(\mathbf{v}^2+1)\alpha^2 - 4\alpha^2\mathbf{v}^2}{8|\alpha - \mathbf{v}^2\sin\alpha\cos\alpha|^3} \to \frac{2\alpha^2(\mathbf{v}_0^2-1)^2}{8\alpha^3(\mathbf{v}_0^2-1)^3} = \frac{1}{4\alpha(\mathbf{v}_0^2-1)} = \frac{1}{4\varepsilon\sqrt{\mathbf{v}_0^2-1}}. \tag{42}$$

As expected the self-force is singular as $\varepsilon \to 0_+ \Rightarrow \Delta \to 0_+$, with $\Delta$ setting the mass scale. In fact the correspondence with mechanical is mass is exact. To see this we restore the constant factor $\mathbf{f}_{norm}$ defined in (34), and note that the speed $|\mathbf{v}|$ therein is required only at the 0$^{th}$ order in $\alpha$, to give the local force in the rotating frame:

$$\mathbf{f}_{local} \to \frac{e^2\omega^2|\mathbf{v}_0|}{2\varepsilon\sqrt{\mathbf{v}_0^2-1}}\hat{\mathbf{x}} = \frac{e^2\omega|\mathbf{v}_0|}{\Delta\sqrt{\mathbf{v}_0^2-1}}\hat{\mathbf{x}}. \tag{43}$$

Compare this with a plausible expression for the ordinary relativistic Newtonian mass-acceleration for a superluminal charge in circular motion at speed constant $|\mathbf{v}_0|$, radian frequency $\omega$, possessing mechanical mass $m_0$:

$$\frac{d}{dt}\frac{m_0|\mathbf{v}_0|}{\sqrt{\mathbf{v}_0^2-1}} = \frac{m_0\mathbf{a}}{\sqrt{\mathbf{v}_0^2-1}} = -\frac{m_0\omega|\mathbf{v}_0|}{\sqrt{\mathbf{v}_0^2-1}}\hat{\mathbf{x}}(t). \tag{44}$$

Cast in the rotating frame of the particle, (44) is the negative of (43) provided one makes the association $m_0 = e^2/\Delta$. The sign difference is because the Newtonian Lorentz equation is $f_{distant} - ma = 0$ whereas the Euler equation for the purely electromagnetic system is $f_{distant} + f_{local} = 0$. That is, if the distant force is considered as `applied', a charge with only electromagnetic mass `responds' with a reaction force whose negative is the traditional mass-acceleration of Newton's second law. It is concluded that $\Delta$ here plays the role of the classical radius of the charge, which, as it tends to zero, causes the electromagnetic mass to tend to infinity. The self-force due to self-action replaces (-1 times) the traditional 'force of inertia'.

## 3.5 The Cerenkov cone condition

Given that the local-force is singular at order $1/\varepsilon$, for the total force to vanish it is necessary that there exist another – canceling - contribution at this order. For some relative phase $\beta$ solving (38) the *distant* force coming from the other particle must be singular at order $1/\varepsilon$ with the correct magnitude to exactly cancel the self-force. Clearly, if it exists, that value of $\beta$ must cause the denominator in (39) to vanish as $\varepsilon \to 0_+$, though, as discussed above, this is a necessary but not sufficient condition to guarantee cancellation of forces at order $1/\varepsilon$. Let $\beta_0 \equiv \beta|_{\varepsilon=0}$ denote the value of $\beta$ in the limit that $\varepsilon \to 0_+ \Rightarrow \Delta \to 0_+ \Rightarrow \alpha = 0_+$. Then for the denominator in (39) to vanish requires

$$\beta_0 + \mathbf{v}_0^2 \sin \beta_0 \cos \beta_0 = 0. \tag{45}$$

To this order (38) gives

$$\beta_0^2 - \mathbf{v}_0^2 \cos^2 \beta_0 = 0. \tag{46}$$

Together these give that the speed is

$$|\mathbf{v}_0| = \operatorname{cosec} \beta_0 \tag{47}$$

where $\beta_0$ is a solution of

$$\beta_0 \tan \beta_0 = -1, \quad \beta_0 > 0. \tag{48}$$

($\beta_0 > 0$ is required by (24)). The first few solutions are given in Table 1. The smallest ($n = 1$) admissible value for $\beta_0$ is 2.798, which corresponds to 320.6° of orbital motion. Figures 2a and 2b illustrate the balance of forces supporting circular motion for this particular mode. In the limit of large speeds the phase and speed approach

$$\beta_0, |\mathbf{v}_0| \to n\pi \tag{49}$$

where $n$ is a positive integer. It will be useful for later to express $\beta_0$ in terms of $|\mathbf{v}_0|$. Writing (46) as

$$\beta_0^2 = \mathbf{v}_0^2 \left(1 - \sin^2 \beta_0\right) \tag{50}$$

and using (47), one obtains

$$\beta_0^2 = \mathbf{v}_0^2 - 1 \tag{51}$$

from which it is inferred that the speed must satisfy

$$\sqrt{\mathbf{v}_0^2 - 1} \tan \sqrt{\mathbf{v}_0^2 - 1} = -1. \tag{52}$$

Every admissible solution of (48) causes the denominator in (39) to vanish to $0^{th}$ order in $\varepsilon$. By virtue of (47) each of these corresponds to a particular orbital speed for both particles. One should be clear though that the complete set of possibilities corresponds to a *single* value for the index *l* in (24). Other values of *l*, if they exist, would correspond to a different intersection by particle 2 of the forward light cone of particle 1 for the *same* parameters $r$, $\omega$, and $|\mathbf{v}_0|$.

Eq. (46) selects the relative phase $\beta_0$ for some fixed speed $|\mathbf{v}_0|$ at which there is electromagnetic contact between the two particles. That is, (46) determines the times at which one particle is on the light cone of the other. For any laboratory time *t* of particle 1, there will always be at least two times (two values for t' in (20)) at which particle 2 is in electromagnetic contact. Independent of the specifics of the motion, at $\Delta = 0$ the light-cone condition (17) is

$$t'_\pm = t \pm |\mathbf{x}_2(t'_\pm) - \mathbf{x}_1(t)|. \tag{53}$$

which gives rise to (37) and (38) in the particular case of circular motion considered here. It will be helpful to regard the coordinate *t* of particle 1 as the fixed present time, and identify with $t'_\pm$ the two times that are the future and past (relative to *t*) times that particle 2 is in electromagnetic contact.

Whereas (37) and (38) and therefore Eq. (53) decides the timing of the electromagnetic interaction, Eq. (45) imposes a constraint on the motion, valid only at that time. Independent of the specifics of the motion and with reference to the denominator of (11), (45) corresponds to the condition

$$u'_2 \circ s_{2,1} = 0 \tag{54}$$

which, in 3+1D form, is

$$t'_\pm - t = \mathbf{v}_2 \cdot (\mathbf{x}_2(t'_\pm) - \mathbf{x}_1(t)). \tag{55}$$

This is the condition that the distant force from particle 2 on particle 1 is singular. As discussed in section 3.4, it is a necessary but not sufficient condition in order that an uncompensated charge deviates from straight-line motion. In the particular case of circular motion discussed above, it has been arranged that the denominator of the distant force vanish, and therefore (54) hold, at *both* times $t'_\pm$, symmetrically distributed either side of *t*. Let us define two unit vectors, each connecting the two particles at the two times of EM contact:

$$\hat{\mathbf{n}}_\pm \equiv \frac{\mathbf{x}_2(t'_\pm) - \mathbf{x}_1(t)}{|\mathbf{x}_2(t'_\pm) - \mathbf{x}_1(t)|}. \tag{56}$$

Then (53) combined with (55) give

$$\mathbf{v}_2 \cdot \hat{\mathbf{n}}_+ = 1, \quad \mathbf{v}_2 \cdot \hat{\mathbf{n}}_- = -1 \tag{57}$$

which gives that at the time of EM contact the component of velocity of particle 2 in the direction of particle 1 – disregarding the sign - is the speed of light. For any superluminal particle the set of such velocities (which forms a double cone in space) is called its Cerenkov cone. More specifically, Eqs. (57) give that particle 2 is moving away from particle 1 at the speed of light at the future time, i.e. as particle 2 crosses the retarded light cone of particle 1, and that particle 2 is moving towards particle 1 at the speed of light at the historical time, i.e. as particle 2 crosses the advanced light cone of particle 1. These conditions will be referred to collectively as the Cerenkov cone condition. In the space-time diagram the conditions (57) mean that particle 2 intersects the (double) light cone of particle 1 at a tangent. Figure 1b illustrates this possibility in 1+1 D. Figures 2a and 2b show how the balance of forces supports circular motion. Figure 3 shows a rendering of the actual motion in 2+1.

## 4. Higher order terms in the Euler equation

### 4.1 Motivation

The calculation of the distant force culminating in (47) and (48) enforces the Cerenkov cone condition discussed in section 3.7. Yet although they guarantee that the distant force is singular, Eqs. (47) and (48) are not sufficient to guarantee that the distant force cancels the self force, for which it is necessary to expand to higher orders. In the particular case of circular motion, it will turn out that higher order corrections will play no direct role in determination of the allowed motion; once the speed have been chosen in accord with (47) and (48) one has in hand a full description of the motion of both particles subject to the prior constraint of concentric circular motion. That is, the outcome of this calculation can have no impact on the speed and relative phase (at the time of EM contact) in the limit $\Delta = 0$. Nonetheless, it is important to show that it is mathematically feasible to *completely* cancel the self-force with the local force, without, say, introducing imaginary quantities. Also, it will turn out that the higher order terms computed below will be required to compute the energy and the angular momentum of the system.

### 4.2 Series solution of the light-cone conditions

We require the simultaneous solution of Eqs. (37), (38) and (40) (with definitions (36), and (39)) for the three unknowns $\alpha$, $\beta$, $\mathbf{v}^2$, as $\varepsilon \to 0_+$. Noting that Eq. (42) gives $f_{local} \sim 1/\varepsilon$, it follows, in order for the total force vanish, that the denominator of the distant force (39) must be proportional to $\varepsilon$ (unless perhaps the numerator vanishes at $\varepsilon = 0$, which will be seen not to be the case). From the form of the denominator, $\left| \beta + \mathbf{v}^2 \sin \beta \cos \beta \right|^3$, it is deduced that $\delta \beta \sim \varepsilon^{1/3}$ at the lowest order and therefore that the quantities $\alpha, \beta, \mathbf{v}^2$ can be expanded as a series in $\varepsilon^{1/3}$. Therefore let us define $\rho = \varepsilon^{1/3}$ (there is just one real root) and write

$$\alpha = \sum_{n=0} \alpha_n \rho^n, \quad \beta = \sum_{n=0} \beta_n \rho^n \tag{58}$$

where $n$ is an integer and the coefficients are constants. It will be convenient to re-write the modified light-cone conditions (37) and (38) using $\rho = \varepsilon^{1/3}$ as

$$\left(\alpha^2 + \rho^6\right)\cos^2\beta - \left(\beta^2 + \sigma_s\rho^6\right)\sin^2\alpha = 0, \tag{59}$$

and

$$\mathbf{v}^2 = \frac{\alpha^2 + \rho^6}{\sin^2\alpha}. \tag{60}$$

The problem may now be stated as the determination of the coefficients $\alpha_n, \beta_n, \gamma_n$ from the simultaneous equations (58), (59) and (60), and (40) (using (36) and (39)).

First we solve for the $\{\alpha_i\}$ in terms of the $\{\beta_i\}$. Substitution of (58) into (59) and collecting terms, one obtains at $0^{\text{th}}$ order

$$\alpha_0^2 \cos^2\beta_0 - \beta_0^2 \sin^2\alpha_0 = 0. \tag{61}$$

Since it has been already deduced that $\beta_0$ satisfies (48), plus that $\alpha \to 0_+$ as $\rho, \varepsilon \to 0_+$, it follows that the above requires that $\alpha_0 = 0$, as expected. At the $1^{\text{st}}$ order the coefficient of $\rho$ in (59) gives

$$\left(-2\alpha_0^2\beta_1\cos\beta_0\sin\beta_0 + 2\alpha_0\alpha_1\cos^2\beta_0 - 2\beta_0^2\alpha_1\cos\alpha_0\sin\alpha_0 - 2\beta_0\beta_1\sin^2\alpha_0\right)\rho = 0 \tag{62}$$

which is clearly already satisfied at $\alpha_0 = 0$. One proceeds likewise, setting the coefficient of each term in the power series for $\rho$ to zero. One can use (48) to replace all appearances of the trig terms with irrational functions of $\beta_0$:

$$\cos\beta_0 = \frac{\beta_0}{\sqrt{1+\beta_0^2}}, \quad \sin\beta_0 = -\frac{1}{\sqrt{1+\beta_0^2}}. \tag{63}$$

One eventually obtains

$$\alpha = \frac{1}{\beta_0}\rho^3 - \frac{\beta_1^2\left(1+\beta_0^2\right)}{2\beta_0^3}\rho^5 + \frac{\beta_1\left(1+\beta_0^2\right)\left(\beta_1^2 - 3\beta_0\beta_2\right)}{3\beta_0^4}\rho^6 \\ + \frac{\left(1+\beta_0^2\right)\left(\beta_1^4\left(1+\beta_0^2\right) + 24\beta_0\beta_1\left(\beta_1\beta_2 - \beta_0\beta_3\right) - 12\beta_0^2\beta_2^2\right)}{24\beta_0^5}\rho^7 + O(\rho^8). \tag{64}$$

Using (64) one can now express $\mathbf{v}^2$ as given by (60) in terms of the $\{\beta_i\}$. Expanding and collecting terms one finds

$$\mathbf{v}^2 = 1 + \beta_0^2 + \beta_1^2\left(1+\beta_0^2\right)\rho^2 + \frac{2\beta_1\left(1+\beta_0^2\right)\left(3\beta_0\beta_2 - \beta_1^2\right)}{3\beta_0}\rho^3$$
$$+ \frac{\left(1+\beta_0^2\right)\left(2\beta_1^4\left(1+\beta_0^2\right) + 6\beta_0\beta_1\left(\beta_0\beta_3 - \beta_1\beta_2\right) + 3\beta_0^2\beta_2^2\right)}{3\beta_0^2}\rho^4 + O\left(\rho^5\right). \tag{65}$$

### 4.3 Series solution of the Euler equation

Using (64) and (65), and the expansion (58) for $\beta$, it is now possible to express the forces entirely in terms of the $\{\beta_i\}$. A series expansion of the Euler equation (40) in $\rho$ will then fix the $\{\beta_i\}$, $i > 0$ in terms of $\beta_0$. One finds for the local force, (36), that

$$\tilde{f}_{local} = \frac{1}{4\beta_0}\left(\frac{1}{\rho^3} - \left(1+\beta_0^2\right)\frac{\beta_1^2}{2\beta_0^2\rho} + O\left(\rho^0\right)\right). \tag{66}$$

Putting (64) and (65) and the expansion (58) for $\beta$ into the expression for the distant force (39) gives

$$\tilde{f}_{distant} = \frac{\text{sgn}(e_1 e_2)}{4\left(1+\beta_0^2\right)\left|\beta_1^3\right|}\left(\frac{1}{\rho^3} + \frac{\beta_1^2 - 3\beta_0\beta_2}{\beta_0\beta_1\rho^2} + \frac{2\beta_1^4 - 2\beta_0\beta_1^2\beta_2 - 3\beta_0^2\beta_1\beta_3 + 6\beta_0^2\beta_2^2}{\beta_0^2\beta_1^2\rho} + O\left(\rho^0\right)\right). \tag{67}$$

It may be observed that $\sigma_s$ does not appear in either of the above forces. This is because $\alpha$ is non-zero only at the third order, so that terms involving $\sigma_s$ first appear in the expansion of (59) at the 12[th] order, which is too high to contribute to the singular components of the force.

The total force must be set to zero by solving (40) using (66) and (67) by setting the coefficient to zero of each of the Taylor-Laurent series terms in $\rho$ up to and including the term at order $\rho^0$. From (66) and (67) the dominant term at order $\rho^{-3}$ gives

$$\frac{\text{sgn}(e_1 e_2)}{\left(1+\beta_0^2\right)\left|\beta_1^3\right|} + \frac{1}{\beta_0} = 0. \tag{68}$$

One sees immediately that $\text{sgn}(e_1 e_2) = -1$, i.e. the two charges are of opposite sign – as expected. Given that this is the case, one then has

$$\beta_1 = \sigma_{\beta 1}\left(\frac{\beta_0}{1+\beta_0^2}\right)^{1/3} \tag{69}$$

where $\sigma_{\beta 1}$ can be either of $\pm 1$. Proceeding likewise for each term of the Taylor-Laurent series, one eventually obtains

$$\beta = \beta_0 + \sigma_{\beta 1}\left(\frac{\beta_0}{1+\beta_0^2}\right)^{1/3}\rho + \frac{1}{3\beta_0}\left(\frac{\beta_0}{1+\beta_0^2}\right)^{2/3}\rho^2 + O\left(\rho^3\right). \tag{70}$$

The $\beta_i$ implied in the above (i.e. as defined in (58)) can now be substituted into (64) and (65) so that $\alpha$ and $\mathbf{v}^2$ can now be regarded as functions of $\rho$ and $\beta_0$. However, it is preferred here to use (51) to express everything in terms of $\mathbf{v}_0^2$ (the square of the speed in the limit that $\rho = 0$), since this is perhaps more readily regarded as an observable property of the two-particle system. In that case the series (64) for $\alpha$ becomes

$$\alpha = \frac{\rho^3}{\sqrt{\mathbf{v}_0^2 - 1}} - \frac{|\mathbf{v}_0|^{2/3}}{2(\mathbf{v}_0^2 - 1)^{7/6}}\rho^5 - \frac{9\mathbf{v}_0^2 + 28}{72|\mathbf{v}_0|^{2/3}(\mathbf{v}_0^2 - 1)^{11/6}}\rho^7 + O(\rho^9) \tag{71}$$

the first term of which is in agreement with (41). The series for $\beta$ is

$$\beta = \sqrt{\mathbf{v}_0^2 - 1} + \sigma_{\beta 1}\frac{(\mathbf{v}_0^2 - 1)^{1/6}}{|\mathbf{v}_0|^{2/3}}\rho + \frac{1}{3|\mathbf{v}_0|^{4/3}(\mathbf{v}_0^2 - 1)^{1/6}}\rho^2 + O(\rho^3) \tag{72}$$

and the series (65) becomes

$$\mathbf{v}^2 = \mathbf{v}_0^2 + |\mathbf{v}_0|^{2/3}(\mathbf{v}_0^2 - 1)^{1/3}\rho^2 + \frac{9\mathbf{v}_0^2 + 7}{9|\mathbf{v}_0|^{2/3}(\mathbf{v}_0^2 - 1)^{1/3}}\rho^4 + O(\rho^6). \tag{73}$$

We can now investigate how each of the terms in the expansion of either the local or the distant force depends on the speed (because the sum is zero, one is just the negative of the other). Examination of (66) and (67) immediately reveals that the coefficient of $\rho^{-2}$ must be zero because a term of that order appears in only one of the two forces. There remains only

$$\tilde{f}_{local} = -\tilde{f}_{dist} = \frac{1}{4\sqrt{\mathbf{v}_0^2 - 1}}\frac{1}{\rho^3} - \frac{|\mathbf{v}_0|^{2/3}}{8(\mathbf{v}_0^2 - 1)^{7/6}}\frac{1}{\rho} + O(\rho^0). \tag{74}$$

To restore the units to the forces we recall $\rho = \varepsilon^{1/3} = (\omega\Delta/2)^{1/3}$ and note also that the normalizing force (34) is itself a function of $\rho$ via $|\mathbf{v}|$, an expansion for which can be obtained from (73):

$$\mathbf{f}_{norm} \equiv 2e^2\omega^2\hat{\mathbf{x}}|\mathbf{v}| = 2e^2\omega^2\hat{\mathbf{x}}\left(|\mathbf{v}_0| + \frac{1}{2}\left(\frac{\mathbf{v}_0^2 - 1}{|\mathbf{v}_0|}\right)^{1/3}\rho^2 + O(\rho^4)\right)$$

$$= 2e^2\omega^2\hat{\mathbf{x}}\left(|\mathbf{v}_0| + \left(\frac{\omega^2\Delta^2(\mathbf{v}_0^2 - 1)}{32|\mathbf{v}_0|}\right)^{1/3} + O(\Delta^{4/3})\right). \tag{75}$$

Multiplying (74) by (75) gives

$$\mathbf{f}_{local} = -\mathbf{f}_{dist} = \frac{e^2\omega|\mathbf{v}_0|\hat{\mathbf{x}}}{\Delta\sqrt{\mathbf{v}_0^2 - 1}} - \frac{e^2\omega^{5/3}\hat{\mathbf{x}}}{2^{5/3}\Delta^{1/3}|\mathbf{v}_0|^{1/3}(\mathbf{v}_0^2 - 1)^{7/6}} + O(\Delta^0). \tag{76}$$

(In fact there are no contributions in (76) to the force at order $\Delta^0$ arising from the foregoing analysis; the first non-singular contribution turns out to be at order $\Delta^{1/3}$. There are, however, contributions at $O(\Delta^0)$ from the finite interactions ignored in this analysis, as discussed above.) As required, the first term is in agreement with (43). Though weaker than the first, the second term is also singular, and therefore likewise requires nullification from balance of local and distance forces.

There were no problems encountered in the above in assigning real values to the expansion coefficients in the series (58) in order that the singular parts of the total force vanish. This confirms that the circular motion of the two charges (18) does indeed satisfy the singular parts of the Euler equations for the action (3) in the limit that $\Delta = 0$ - provided the speed is a solution of (52). The non-singular parts, in particular the total force at $O(\Delta^0)$ can similarly be nullified through appropriate choice of the coefficients of the higher order terms on the expansions (58). As discussed above, this is not demonstrated here because of space limitations and also because it has no bearing on the quantities of interest. The main point, that the charges interact to produce non-trivial motions – despite the infinite electromagnetic masses, is sufficiently demonstrated by resolving just the singular parts of the force.

## 5. Energy and Angular Momentum

### 5.1 Energy of pair in circular motion

In this section the local and then the distant contributions to the energy are first computed separately. The total energy is then given as the sum. In the case that the motion is given by (18) and the scalar products are (27), the energy is

$$H = \frac{1}{2}\sum_{j,k}\sum_{roots} \left\langle \frac{e_j e_k \omega}{|\theta - \sigma_{k,j}\mathbf{v}^2 \sin\theta|^3} \begin{pmatrix} 2(\theta - \sigma_{k,j}\mathbf{v}^2 \sin\theta)^2 - 2(\theta - \sigma_{k,j}\mathbf{v}^2 \sin\theta)(1 - \sigma_{k,j}\mathbf{v}^2 \cos\theta)\theta \\ -\theta^2(\theta - \sigma_{k,j}\mathbf{v}^2 \sin\theta)\sigma_{k,j}\mathbf{v}^2 \sin\theta + \theta^2(1 - \sigma_{k,j}\mathbf{v}^2 \cos\theta)^2 \end{pmatrix} \right\rangle $$

$$= \sum_{j,k}\sum_{roots>0} \frac{e_j e_k \omega}{|\theta - \sigma_{k,j}\mathbf{v}^2 \sin\theta|^3}\left(\mathbf{v}^4(\theta^2 + 2\sin\theta(\sin\theta - \theta\cos\theta)) - \sigma_{k,j}\mathbf{v}^2\theta(\theta^2 + 2)\sin\theta + \theta^2\right) \tag{77}$$

The sum over $\sigma_s$ has been dropped because, as discussed above, both the local and distant interactions can simultaneously satisfy the modified light-cone condition and the Cerenkov cone condition for just one of the two values. The roots are now given by (23). The angle brackets have been be omitted because nothing inside depends on $t$. Noting that the whole expression is even in $\theta$, and that (23) is even in $\theta$ and therefore the roots come in opposite signed pairs, the sum over roots generates a factor of 2, and it is understood that in the second expression above one now uses just the positive root of (23).

In accord with the discussion of section 3.4 we distinguish between energy arising from local self-action and any other energy arising from the action of the 'distant' force:

$$H = H_{local} + H_{dist}.\tag{78}$$

The local energy is characterized by $j = k$ and $\theta \to 0_+$ as $\Delta \to 0_+$. Using $\theta = 2\alpha$ and that there are two particles having local contributions from the pairs $(j,k) = (1,1)$ and $(j,k) = (2,2)$, (77) becomes

$$H_{local} = \frac{4e^2\omega}{\left|2\alpha - \mathbf{v}^2 \sin 2\alpha\right|^3}\left(\mathbf{v}^4\left(2\alpha^2 + \sin 2\alpha(\sin 2\alpha - 2\alpha \cos 2\alpha)\right) - \mathbf{v}^2 2\alpha\left(2\alpha^2 + 1\right)\sin 2\alpha + 2\alpha^2\right)\tag{79}$$

where $\alpha$ is the solution of (37). Using now the expansion (71) for $\alpha$ and (73) for $\mathbf{v}^2$, after some algebra one finds

$$H_{local} = \frac{e^2\omega}{\rho^3\sqrt{\mathbf{v}_0^2 - 1}} + \frac{e^2\omega|\mathbf{v}_0|^{2/3}}{2\rho\left(\mathbf{v}_0^2 - 1\right)^{7/6}} + O(\rho^0) = \frac{2e^2}{\Delta\sqrt{\mathbf{v}_0^2 - 1}} + \frac{e^2\omega^{2/3}|\mathbf{v}_0|^{2/3}}{2^{2/3}\Delta^{1/3}\left(\mathbf{v}_0^2 - 1\right)^{7/6}} + O(\Delta^0).\tag{80}$$

Just as for the force, the term at order $\Delta^0$ comes from finite interactions ignored in the foregoing analysis. The first term is just that expected from a pair of particles each of (electromagnetic) rest-mass $e^2/\Delta$, provided one accepts the generalization, already implied in (80), of the traditional $\gamma$-factor to the superluminal domain.

Taking into account the exchange symmetry between the two particles, the interaction energy is two times that of just one of them. Using $\theta = 2\beta$ and that there are two oppositely-charged particles corresponding to the pairs $(j,k) = (1,2)$ and $(j,k) = (2,1)$, both pairs assumed to contribute equally, (77) becomes

$$H_{dist} = -\frac{4e^2\omega}{\left|2\beta + \mathbf{v}^2 \sin 2\beta\right|^3}\left(\mathbf{v}^4\left(2\beta^2 + \sin 2\beta(\sin 2\beta - 2\beta \cos 2\beta)\right) + \mathbf{v}^2 2\beta\left(2\beta^2 + 1\right)\sin 2\beta + 2\beta^2\right)\tag{81}$$

where $\beta$ is the solution of (38). Using now the expansion (72) for $\beta$ and (73) for $\mathbf{v}^2$, one obtains

$$H_{dist} = -\frac{e^2\omega \mathbf{v}_0^2}{\rho^3\sqrt{\mathbf{v}_0^2 - 1}} - \frac{e^2\omega|\mathbf{v}_0|^{2/3}\left(\mathbf{v}_0^2 - 3\right)}{3\rho\left(\mathbf{v}_0^2 - 1\right)^{7/6}} + O(\rho^0) = -\frac{2e^2\mathbf{v}_0^2}{\Delta\sqrt{\mathbf{v}_0^2 - 1}} - \frac{2^{1/3}e^2\omega^{2/3}|\mathbf{v}_0|^{2/3}\left(\mathbf{v}_0^2 - 3\right)}{3\Delta^{1/3}\left(\mathbf{v}_0^2 - 1\right)^{7/6}} + O(\Delta^0).\tag{82}$$

The interaction energy is negative, as it should be.

Adding (80) and (82), the total energy is

$$H = -\frac{2e^2\sqrt{\mathbf{v}_0^2 - 1}}{\Delta} - \frac{e^2\omega^{2/3}|\mathbf{v}_0|^{2/3}}{2^{2/3}\Delta^{1/3}\left(\mathbf{v}_0^2 - 1\right)^{1/6}} + O(\Delta^0).\tag{83}$$

The total energy of the system is negative, confirming that the orbits are binding. The energy is not bounded from below; increasing the speed makes the energy more negative. Since $\mathbf{v}_0^2$ must obey (52), the energy is quantized. Table 2 gives the spectrum for the first few quantum numbers. At high quantum numbers (49) gives that the energy approaches

$$H \to -2e^2 \pi n / \Delta, \tag{84}$$

where $n$ is a positive integer.

Denoting the equivalent rest mass of each particle by $m = e^2/\Delta$ it may be noticed that to first order in the mass (83) appears to be a super-luminal generalization of the relativistic Virial Theorem [23]

$$H = \sum_j m_j \left\langle \sqrt{1 - \mathbf{v}_j^2} \right\rangle. \tag{85}$$

However, a formal relation between the two has not been established here; that would require demonstration of (83) independent of the details of the motion.

In this purely electromagnetic theory the coefficient of the action (3) is entirely arbitrary. A coefficient proportional to $\Delta$, say, would in effect be a form of renormalization, which would leave the energy (83) finite and the angular momentum (96) zero. In any case, the absolute values of the energy and angular momentum must be regarded as arbitrary. Of course, the fact of their quantization is unaffected by this arbitrariness, since the excited states can be expressed as dimensionless ratios relative to the ground state. For example, upon introducing an extra subscript to denote the quantum number, so that, for example, the $n^{th}$ solution to (52) is labeled $\mathbf{v}_{0,n}^2$, then (83) gives

$$\lim_{\Delta \to 0_+} \frac{H_n}{H_0} = \sqrt{\frac{\mathbf{v}_{0,n}^2 - 1}{\mathbf{v}_{0,0}^2 - 1}} \xrightarrow[\text{large } n]{} \frac{n\pi}{\sqrt{\mathbf{v}_{0,0}^2 - 1}} \approx 1.301 n \tag{86}$$

where we used $|\mathbf{v}_{0,0}| \approx 2.614$.

### 5.2 Angular momentum of pair in circular motion

In this section the local and then the distant contributions to the angular momentum are first computed separately. The total angular momentum is then given as the sum. First we simplify the third term in large parentheses in (A77) using the scalar products for the dual circular motion (27):

$$\left(\mathbf{u}'_k \circ \mathbf{s}_{k,j}\right)\left(\mathbf{u}'_k \circ \mathbf{u}_j + (t'-t)\mathbf{a}'_k \circ \mathbf{u}_j\right) - (t'-t)\left(\mathbf{u}'^2_k + \mathbf{a}'_k \circ \mathbf{s}_{k,j}\right)\left(\mathbf{u}'_k \circ \mathbf{u}_j\right)$$
$$= \frac{1}{\omega}\left(\theta - \sigma_{k,j}\mathbf{v}^2 \sin\theta\right)\left(1 - \sigma_{k,j}\mathbf{v}^2 \cos\theta + \frac{\theta}{\omega}\sigma_{k,j}\mathbf{v}^2 \omega \sin\theta\right) - \frac{\theta}{\omega}\left(1 - \sigma_{k,j}\mathbf{v}^2 \cos\theta\right)^2. \quad (87)$$
$$= \frac{\mathbf{v}^2}{\omega}\left(\mathbf{v}^2 (\sin\theta \cos\theta - \theta) + \sigma_{k,j}\left((\theta^2 - 1)\sin\theta + \theta \cos\theta\right)\right)$$

From the definitions (18) and (19) the vector cross products are:

$$\begin{aligned}
\mathbf{x}'_k \times \mathbf{x}_j &= -\hat{\mathbf{z}}\frac{\mathbf{v}^2}{\omega^2}\sin\left(\omega(t'-t) + \phi_k - \phi_j\right) = -\sigma_{k,j}\hat{\mathbf{z}}\frac{\mathbf{v}^2}{\omega^2}\sin\theta \\
\mathbf{v}'_k \times \mathbf{v}_j &= -\hat{\mathbf{z}}\mathbf{v}^2 \sin\left(\omega(t'-t) + \phi_k - \phi_j\right) = -\sigma_{k,j}\hat{\mathbf{z}}\mathbf{v}^2 \sin\theta \\
\mathbf{x}'_k \times \mathbf{v}_j &= \hat{\mathbf{z}}\frac{\mathbf{v}^2}{\omega}\cos\left(\omega(t'-t) + \phi_k - \phi_j\right) = \sigma_{k,j}\hat{\mathbf{z}}\frac{\mathbf{v}^2}{\omega}\cos\theta \\
\mathbf{v}'_k \times \mathbf{x}_j &= -\hat{\mathbf{z}}\frac{\mathbf{v}^2}{\omega}\cos\left(\omega(t'-t) + \phi_k - \phi_j\right) = -\sigma_{k,j}\hat{\mathbf{z}}\frac{\mathbf{v}^2}{\omega}\cos\theta
\end{aligned} \quad (88)$$

Then the first vector expression in (A77) is

$$\left(\mathbf{x}_j \times \mathbf{v}'_k + \left(\mathbf{x}'_k + (t'-t)\mathbf{v}'_k\right) \times \mathbf{v}_j\right) = \sigma_{k,j}\hat{\mathbf{z}}\frac{\mathbf{v}^2}{\omega}\cos\theta + \sigma_{k,j}\hat{\mathbf{z}}\frac{\mathbf{v}^2}{\omega}\cos\theta - \frac{\theta}{\omega}\sigma_{k,j}\hat{\mathbf{z}}\mathbf{v}^2 \sin\theta$$
$$= \sigma_{k,j}\hat{\mathbf{z}}\frac{\mathbf{v}^2}{\omega}(2\cos\theta - \theta\sin\theta) \quad (89)$$

Putting (87)-(89) in (A77) one obtains the angular momentum specific to the dual circular motion

$$\mathbf{L} = \sum_{j,k}\sum_{\text{roots}} \frac{e_j e_k}{2} \frac{\hat{\mathbf{z}}\sigma_{k,j}\mathbf{v}^2}{\left|\theta - \sigma_{k,j}\mathbf{v}^2 \sin\theta\right|^3}
\begin{pmatrix}
\left(\theta - \sigma_{k,j}\mathbf{v}^2 \sin\theta\right)^2 (2\cos\theta - \theta\sin\theta) \\
-\left(\theta - \sigma_{k,j}\mathbf{v}^2 \sin\theta\right)\left(1 - \sigma_{k,j}\mathbf{v}^2 \cos\theta\right)\theta\cos\theta \\
-\mathbf{v}^2\left(\mathbf{v}^2(\sin\theta\cos\theta - \theta) + \sigma_{k,j}\left((\theta^2-1)\sin\theta + \theta\cos\theta\right)\right)\sin\theta
\end{pmatrix}$$
$$= \sum_{j,k}\sum_{\text{roots}>0} \frac{\hat{\mathbf{z}}\sigma_{k,j}e_j e_k \mathbf{v}^2}{\left|\theta - \sigma_{k,j}\mathbf{v}^2 \sin\theta\right|^3}
\begin{pmatrix}
\mathbf{v}^4 \sin^2\theta \cos\theta \\
+\sigma_{k,j}\mathbf{v}^2\left(\sin^2\theta + \theta^2 - 4\theta\sin\theta\cos\theta\right) \\
+\theta^2(\cos\theta - \theta\sin\theta)
\end{pmatrix} \quad (90)$$

In accord with previous discussions the sum over $\sigma_s$ has been dropped. The angle brackets have been be omitted because nothing inside depends on $t$. Given that the whole expression is even in $\theta$ and that (23) is even in $\theta$ - and therefore the roots come in opposite signed pairs – the sum over roots has been replaced with a factor of 2 and the positive root of (23) has been stipulated.

Proceeding as for the energy let

$$\mathbf{L} = \mathbf{L}_{local} + \mathbf{L}_{dist} \quad (91)$$

where the local contribution is characterized by $j = k$ and $\theta \to 0_+$ as $\Delta \to 0_+$. Using $\theta = 2\alpha$ and that there are two particles corresponding to the pairs $(j,k) = (1,1)$ and $(j,k) = (2,2)$, (90) becomes

$$\mathbf{L}_{local} = -\hat{\mathbf{z}} \frac{2e^2\mathbf{v}^2}{|2\alpha - \mathbf{v}^2\sin 2\alpha|^3} \begin{pmatrix} \mathbf{v}^4 \sin^2 2\alpha \cos 2\alpha + \mathbf{v}^2\left(\sin^2 2\alpha + 4\alpha^2 - 8\sin 2\alpha \cos 2\alpha\right) \\ +4\alpha^2\left(\cos 2\alpha - 2\alpha \sin 2\alpha\right) \end{pmatrix} \qquad (92)$$

where $\alpha$ is the solution of (37). Using now the expansion (71) for $\alpha$ and (73) for $\mathbf{v}^2$, this is

$$\begin{aligned}\mathbf{L}_{local} &= \hat{\mathbf{z}} e^2 \left( \frac{\mathbf{v}_0^2}{\rho^3\sqrt{\mathbf{v}_0^2-1}} + \frac{\mathbf{v}_0^{2/3}\left(\mathbf{v}_0^2-2\right)}{2\rho\left(\mathbf{v}_0^2-1\right)^{7/6}} \right) + O(\rho^0) \\ &= \hat{\mathbf{z}} e^2 \left( \frac{2\mathbf{v}_0^2}{\omega\Delta\sqrt{\mathbf{v}_0^2-1}} + \frac{\mathbf{v}_0^{2/3}\left(\mathbf{v}_0^2-2\right)}{2^{2/3}\omega^{1/3}\Delta^{1/3}\left(\mathbf{v}_0^2-1\right)^{7/6}} \right) + O(\Delta^0)\end{aligned} \qquad (93)$$

In computing the distant contribution to the angular momentum we use $\theta = 2\beta$ and that there are two oppositely-charged particles corresponding to the pairs $(j,k) = (1,2)$ and $(j,k) = (2,1)$, and that both pairs contribute equally. Eq. (90) then gives

$$\mathbf{L}_{dist} = \hat{\mathbf{z}} \frac{2e^2\mathbf{v}^2}{|2\beta + \mathbf{v}^2\sin 2\beta|^3} \left(\mathbf{v}^4 \sin^2 2\beta \cos 2\beta + \mathbf{v}^2\left(\sin^2 2\beta + 4\beta^2\right) + 4\beta^2\left(\cos 2\beta + 2\beta \sin 2\beta\right)\right) \qquad (94)$$

where $\beta$ is the solution of (38). Using now the expansion (72) for $\beta$ and (73) for $\mathbf{v}^2$ this is

$$\begin{aligned}\mathbf{L}_{dist} &= -\frac{\hat{\mathbf{z}} e^2 \mathbf{v}_0^2}{\rho^3\sqrt{\mathbf{v}_0^2-1}} + \frac{\hat{\mathbf{z}} e^2 \mathbf{v}_0^{8/3}}{2\rho\left(\mathbf{v}_0^2-1\right)^{7/6}} + O(\rho^0) \\ &= -\frac{2\hat{\mathbf{z}} e^2 \mathbf{v}_0^2}{\omega\Delta\sqrt{\mathbf{v}_0^2-1}} + \frac{\hat{\mathbf{z}} e^2 \mathbf{v}_0^{8/3}}{2^{2/3}\omega^{1/3}\Delta^{1/3}\left(\mathbf{v}_0^2-1\right)^{7/6}} + O(\Delta^0)\end{aligned} \qquad (95)$$

Comparing (93) and (95) one sees that the leading order parts of the angular momentum cancel, with the result that the total momentum is singular only to order $1/\rho$:

$$\mathbf{L} = \hat{\mathbf{z}} \frac{e^2 \mathbf{v}_0^{2/3}}{\rho\left(\mathbf{v}_0^2-1\right)^{1/6}} + O(\rho^0) = \hat{\mathbf{z}} \frac{2^{1/3} e^2 \mathbf{v}_0^{2/3}}{\omega^{1/3}\Delta^{1/3}\left(\mathbf{v}_0^2-1\right)^{1/6}} + O(\Delta^0). \qquad (96)$$

The angular momentum is quantized by virtue of (52). At high quantum numbers, (49) gives that the magnitude of the angular momentum approaches

$$|\mathbf{L}| \to e^2 \left(\frac{2n\pi}{\omega\Delta}\right)^{1/3} \qquad (97)$$

where $n$ is a positive integer.

# 6. Discussion

**6.1 Cerenkov Radiation**

The advantage of the direct-action analysis is that without additional matter there can be no radiation from the circular orbits, i.e. even though there is acceleration. That is, the analysis above is exact for a 2-particle universe. This is in contrast to the Maxwell theory in which there will be (retarded) radiation due to the acceleration of the sources. With this caveat it was stated in the introduction that the technique of non-compensation, and in particular the analysis of the dual circular motion, could also be performed in a Maxwell framework because the theories are otherwise the same.

There is an additional reason, however, for preferring the direct-action theory that was not mentioned in the introduction. It is widely held that *uniformly* moving charges at $v > c$ will radiate into a vacuum. Such radiation was predicted by Sommerfeld [24] in 1904. There were no known physical manifestations, and presumably were not expected to be any - at least not due to ordinary electrons whose speed could not exceed $c$ - until 1934 when Cerenkov and Vavilov observed luminescence produced by relativistic electrons passing though a medium. The phenomenon was explained in 1958 by Cerenkov, Tamm and Frank [25] as an instance of the effect predicted by Sommerfeld, but where the condition for radiation was reformulated as the requirement that the charged particle move faster than the speed of light in the local material. See [26] for a supporting calculation. Wimmel & Jones [27,28] later re-analyzed the production of Cerenkov radiation, replacing Sommerfeld's model of a rigid finite-sized electron (introduced to provide a cutoff for the radiation) with a relativistically-correct variable form, to achieve a finite relativistically-invariant expression for the power loss. But there were features of the result that led Wimmel to conclude that something was wrong with the whole approach. There is a detailed discussion of this issue and a suggested solution in the recent book by Fayngold [29].

However, there are some - for example [30-32] - who hold that Sommerfeld's original analysis applies only to electrons moving through a dense medium, and does not apply to motion in a vacuum. In particular, Recami and Mignani [30] make an extended ($v > c$) Lorentz transformation [33] of the behavior of ordinary charged tardyons to deduce that Cerenkov radiation from charged tachyons appears only if traveling with speed less than that of light in a medium, i.e. when $c/n > v > c$. (The refractive index $n$ is less than 1 in tachyonic material.)

An additional advantage, therefore, of adopting the direct action over the Maxwell theory is that it avoids having to take sides on this contested issue. Just as for ordinary 'acceleration radiation', in the direct-action theory there can be no Cerenkov radiation of energy without the presence of additional absorbers, regardless of which position on the issue is correct.

## 6.2 Real versus imaginary mass

Commonly it assumed that superluminal speeds connote an imaginary value for the mass in order that the classical mechanical action $I_{mech} = -m \int \sqrt{dx^2}$ remains real [29]. A consequence is that one is forced to regard tardyonic ($v < c$) and tachyonic ($v > c$) charged matter as different species, each making their own distinct contribution to the action:

$$I_{mech} = -\sum_{i=1}^{N} m_i \int \sqrt{dx_i^2} - \sum_{i=1}^{N^*} m_i^* \int \sqrt{-dx_i^2} \tag{98}$$

where $m$ and $m^*$ are tardyon and tachyon masses respectively. This is consistent with the Himalayan analogy deployed by Sudarshan in defense of tachyons [34]. By contrast it may be inferred from the analysis above that a mass of purely electromagnetic origin contributes an action

$$I_{self} = -\frac{e^2}{\Delta} \int \sqrt{|dx_i^2|} \tag{99}$$

valid therefore for all speeds. Of course there is nothing in the above to suggest the possibility of different masses, or even of the emergence of a finite mass scale; these are separate issues that require treatment elsewhere. The important point here is that a single action now covers the full domain of speeds, with no mention necessary of imaginary mass.

It appears that in un-compensated EM a single species of charge has the capacity to move with speeds both less than and greater than light. It should be pointed out though that the possibility of null motion is not covered by this conclusion and has yet to be investigated. (Initially it might appear that such a possibility is excluded because one always expects a self-force of the form (43), and therefore an infinite contribution from the gamma factor as $v \to c$. But a more careful analysis shows that if $v = c$ is stipulated at the outset, a different Taylor-Laurent series expansion in $\rho$ of the local force exists, i.e., having finite coefficients, and therefore could potentially be cancelled by a distant force of the same order.) Only if $v = c$ is permitted is there a possibility of a single charge (an unbroken world line) exploring both regimes.

## 6.3 Non-locality

It is clear from (10) and (11) that $u'_k \circ s_{k,j} = 0$ with $s_{k,j}^2 = 0$ is the Lorentz-invariant condition that there exists a singular force between particles (more exactly, events) $k$ and $j$. If the speed of both particles is less than $c$ then this condition is satisfied only at $\mathbf{x}_k = \mathbf{x}_j$ I.E., when the two particles are coincident, in which case the singular force is the Coulomb force at zero separation. For the purposes of computing singular forces, therefore, $u'_k \circ s_{k,j} = 0$ with $s_{k,j}^2 = 0$ is the Lorentz-Invariant generalization of the zero separation condition, from which it follows that a superluminal charge should be regarded (for these purposes) as an extended object. More specifically, for each 4-point, $z(\lambda)$ say, on the world line, a source is 'effectively' extended throughout the surface whose points $\{x(\lambda)\}$ simultaneously satisfy the two equations

$$(z-x)^2 = 0, \quad \frac{dz}{d\lambda} \circ (z-x) = 0. \tag{100}$$

For a single value of $\lambda$ the surface $\{x(\lambda)\}$ has co-dimension 2 (i.e. a 2-surface in 3+1D), so that the 1 dimensional world-line generates an extended source of co-dimension 1 (a volume in 3+1D). This extension is in space-time, not in space alone. For example, a uniformly moving source with speed $v > c$ generates a (double) Cerenkov cone – originating from each point of the world line - whose angle $\theta$ between the surface and the axis of symmetry satisfies $v\cos\theta = c$. This double cone exists for all time, and therefore has 3 dimensions in 3+1. The additional constraint that the 'extended source' lie on the (double) light cone then selects a moving cross-section through the double cone that is a pair of expanding circles of radius $\rho = t\sin\theta$ running away from the source-point at speed $c$, i.e. with $r = \pm t$. A pair of expanding circles is generated by every point indexed by $\lambda$ on the world line; they are the effective extension of the charge – at least insofar as they determine a surface upon which a test charge will experience a singular electromagnetic force. Due to the effective space-time extension of a superluminal charge, at every point in time on the dual circular motion analyzed above each charge feels an infinite force from its partner from two places at once, as illustrated in Figure 2. (It may additionally feel the influence of any number of finite forces, though these do not arise from the effective extension but from zero-dimensional points on the world line.) Each instance of the infinite force arises from those points where the world-line of one charge source intersects the effective extension (of co-dimension 1) of the other source.

## Conclusion

From the analysis of two particles in a positronium-like configuration it has been demonstrated that there exists a non-trivial physics of classical direct-action electromagnetism of point sources with mechanically un-compensated electromagnetic mass, even though those masses are infinite. In the absence of mechanical mass adequate infinite forces have been found to arise from distant charges in superluminal motion. The forces can be regarded as deriving from superluminally-induced change in dimensionality of a source from a 1D world line to a hyper-surface in 3+1D. In the particular case analyzed it was found that the adequate infinite forces exist only for discrete values of the parameters of the motion, resulting in quantized values for the total energy and angular momentum.

# Appendix A

# Force, energy and angular momentum of a system of particles governed by a two-time action

## A.1 Introduction

In classical mechanics Noether's theorem is usually applied to an action has just one time variable and contains only first derivatives:

$$I = \int dt L(t;q,\dot{q}),  \tag{A1}$$

where $q$ can have any number of components. Barut [35] gives a generalization to an arbitrary number of derivatives. However, published derivations of conserved quantities associated with a classical *two-time* Lagrangian,

$$I = \int d\lambda \int d\lambda' J(\lambda' - \lambda; q(\lambda), q(\lambda'), \dot{q}(\lambda), \dot{q}(\lambda')),  \tag{A2}$$

have not been found. The solution to this deficit adopted here is to cast (A2) in the one-time form (A1) by introducing auxiliary degrees of freedom subject to constraints, as described in section A.2, and then apply Noether's theorem to that one-time action. In section A.3 it is demonstrated that the new 'alternative' one-time form gives the same Euler equation as obtained for the original two-time form. In that analysis $q$ is a 4-component vector that is the locus of the particle world-line. The corresponding Euler equation is valid for trajectories that time-reverse. If it is known a priori that there are no time reversals, then a very similar analysis can be performed restricting $q$ to three degrees of freedom, the results of which are simply stated in section A.4. The resulting one-time action is the starting point for computing the energy in section A.5 and in section A.6 the energy is given for the particular case that the original two-time action is that of direct action EM. In section A.7 is computed the angular momentum for the alternative action of section 4 and section A.8 gives the angular momentum for the particular case that the original two-time action is that of direct action EM.

## A.2 Conversion from two-time to one-time form

In this appendix we will consider the general case of a two-time action that can be written in the form

$$I = \int d\lambda \int d\lambda' J(\lambda' - \lambda, x_1, x_2, ..., u_1, u_2, ..., x'_1, x'_2, ..., u'_1, u'_2, ...) = \int d\lambda L_J  \tag{A3}$$

where

$$L_J \equiv \int d\kappa\, J(\kappa, x_1, x_2, ..., u_1, u_2, ..., x_1^+, x_2^+, ..., u_1^+, u_2^+, ...)  \tag{A4}$$

and where the following shorthand has been used for the 4-vectors $x$ and $u$

$$x_j \equiv x_j(\lambda), \quad u_j \equiv \frac{dx_j(\lambda)}{d\lambda}, \quad x'_j \equiv x_j(\lambda'), \quad u'_j \equiv \frac{dx_j(\lambda')}{d\lambda'}, \quad x_j^+ \equiv x_j(\lambda+\kappa), \quad u_j^+ \equiv u_j(\lambda+\kappa). \tag{A5}$$

Note the form (A3) precludes a more general possibility wherein the kernel depends explicitly on both times independently; it depends explicitly only on the difference between the two times. Clearly, the direct-action (3) is in this form. In particular,

$$J = -\sum_{j,k}\sum_{\sigma_s=\pm 1} e_j e_k u_j \circ u_k^+ \delta\left((x_k^+ - x_j)^2 + \sigma\Delta^2\right). \tag{A6}$$

Eq. (A3) can be written as a one-time action by expressing the position and velocity at time $\lambda+\kappa$ in terms of the derivatives of the position at time $\lambda$ as a Taylor series:

$$x_j^+ = \sum_{m=0} \frac{\kappa^m}{m!} x_j^{(m)}, \quad u_j^+ = \sum_{m=0} \frac{\kappa^m}{m!} x_j^{(m+1)}. \tag{A7}$$

Here and subsequently it is assumed that the trajectory is infinitely differentiable, and that series such as (A7) converge. Because we wish to cast (A3) as a one-time action with only first derivatives, we now proceed as if, for each $m$, $x_j^{(m)}$ in (A7) is an independent variable to be varied when extremizing the action (A3), and establish the association between $x_j^{(m)}$ and $(d/d\lambda)^m x_j$ with Lagrange multipliers. In place of (A3) therefore, consider the action $I = \int d\lambda L$ where

$$L = L_{\bar{J}}\left(x_1^{(0)}, x_2^{(0)}, \ldots, x_1^{(1)}, x_2^{(1)}, \ldots, x_1^{(2)}, x_2^{(2)}, \ldots, \ldots\right) + L_{aux}\begin{pmatrix} x_1^{(0)}, x_2^{(0)}, \ldots, x_1^{(1)}, x_2^{(1)}, \ldots, x_1^{(2)}, x_2^{(2)}, \ldots, \ldots, \\ \dot{x}_1^{(0)}, \dot{x}_2^{(0)}, \ldots, \dot{x}_1^{(1)}, \dot{x}_2^{(1)}, \ldots, \dot{x}_1^{(2)}, \dot{x}_2^{(2)}, \ldots, \ldots, \\ p_1^{(0)}, p_2^{(0)}, \ldots, p_1^{(1)}, p_2^{(1)}, \ldots, p_1^{(2)}, p_2^{(2)}, \ldots, \ldots \end{pmatrix}, \tag{A8}$$

where $L_{aux}$ is the auxiliary density

$$L_{aux}(\lambda) = \sum_{j,m} p_j^{(m)} \circ \left(\dot{x}_j^{(m)} - x_j^{(m+1)}\right). \tag{A9}$$

$L_{\bar{J}}$ is the original density $L_J$ but with the replacements

$$x_j \to x_j, \quad u_j \to x_j^{(1)}, \quad x_j^+ \to \sum_m \frac{\kappa^m}{m!} x_j^{(m)}, \quad u_j^+ \to \sum_m \frac{\kappa^m}{m!} x_j^{(m+1)} \tag{A10}$$

so that (A4) becomes

$$L_{\bar{J}} \equiv \int d\kappa \, J\left(\kappa, x_1^{(0)}, x_2^{(0)}, \ldots, x_1^{(1)}, x_2^{(1)}, \ldots, \sum_m \frac{\kappa^m}{m!} x_1^{(m)}, \sum_m \frac{\kappa^m}{m!} x_2^{(m)}, \ldots, \sum_m \frac{\kappa^m}{m!} x_1^{(m+1)}, \sum_m \frac{\kappa^m}{m!} x_2^{(m+1)}, \ldots\right). \tag{A11}$$

It may be observed that $L_{\bar{J}}$ contains no derivatives (at all) and is in one-time form; the functions $x_j^{(m)}$ depend only on $\lambda$. In these terms the particular case (A6) of direct action EM becomes

$$L_{\bar{j}} = -\sum_{j,k} \sum_{\sigma_s = \pm 1} e_j e_k \int d\kappa \sum_m \frac{\kappa^m}{m!} x_j^{(1)} \circ x_k^{(m+1)} \delta\left(\left(\sum_m \frac{\kappa^m}{m!} x_k^{(m)} - x_j^{(0)}\right)^2 + \sigma_s \Delta^2\right). \tag{A12}$$

### A.3 Euler equations

The goal of the following is to show that the action (A8) is equivalent to (A3) in that the resulting Euler equations for the particle trajectories are the same. The Euler equations for (A3) are very simply obtained as follows. The increment in the action due to an increment in the trajectories is:

$$\delta I = \int d\lambda \int d\lambda' \sum_I \left(\delta x_I \circ \frac{\partial}{\partial x_I} + \delta u_I \circ \frac{\partial}{\partial u_I} + \delta x'_I \circ \frac{\partial}{\partial x'_I} + \delta u'_I \circ \frac{\partial}{\partial u'_I}\right) J. \tag{A13}$$

Integrating the 2$^{nd}$ and 4$^{th}$ terms by parts, and assuming that the increments vanish on the boundary (i.e. in the remote $\lambda$-past and remote $\lambda$-future), this is

$$\delta I = \sum_I \int d\lambda \int d\lambda' \left(\delta x_I \circ \left(\frac{\partial}{\partial x_I} - \frac{d}{d\lambda}\frac{\partial}{\partial u_I}\right) + \delta x'_I \circ \left(\frac{\partial}{\partial x'_I} - \frac{d}{d\lambda'}\frac{\partial}{\partial u'_I}\right)\right) J. \tag{A14}$$

Renaming $\lambda$ as $\lambda'$ and $\lambda'$ as $\lambda$ one obtains

$$\delta I = \sum_I \int d\lambda' \int d\lambda\, \delta x_I \circ \left(\frac{\partial}{\partial x_I} - \frac{d}{d\lambda}\frac{\partial}{\partial u_I}\right)\left(1 + \hat{S}(\lambda \leftrightarrow \lambda')\right) J \tag{A15}$$

where $\hat{S}$ operates on $J$ swapping $\lambda$ and $\lambda'$. At an extremum, the increment $\delta I$ vanishes for arbitrary $\delta x_j$. It follows that the Euler equations for (A3) are

$$\int d\lambda' \left(\frac{\partial}{\partial x_I} - \frac{d}{d\lambda}\frac{\partial}{\partial u_I}\right)\left(1 + \hat{S}(\lambda \leftrightarrow \lambda')\right) J = 0 \quad \forall I. \tag{A16}$$

The Euler equations for the system (A8) can be computed as follows. Variation of the auxiliary variables $p_j^{(m)}$ in the auxiliary action (A9) immediately establishes the required relation between $x_j^{(m)}(\lambda)$ and $(d/d\lambda)^m x_j(\lambda)$:

$$\forall k, m \quad \dot{x}_j^{(m)} = x_j^{(m+1)}. \tag{A17}$$

Eq. (A9) also gives that

$$\partial L / \partial \dot{x}_I^{(n)} = p_I^{(n)}. \tag{A18}$$

Both the auxiliary and the original action (A11) participate in giving

$$\partial L / \partial x_I^{(n)} = g_I^{(n)} - p_I^{(n-1)} \Theta_{n-1} \tag{A19}$$

where $g$ is the 4-vector

$$g_I^{(n)} \equiv \int d\kappa \left\{ \frac{\partial J}{\partial x_I} \delta_{n,0} + \frac{\partial J}{\partial u_I} \delta_{n,1} + \frac{\kappa^n}{n!} \frac{\partial J}{\partial x_I^+} + \frac{\kappa^{n-1}}{(n-1)!} \frac{\partial J}{\partial u_I^+} \Theta_{n-1} \right\}. \tag{A20}$$

Here the $\delta$'s are Kronecker symbols, and $\Theta$ is the asymmetric Heaviside step function limited to integer arguments which is zero for negative argument and otherwise one, so that $\Theta_0 = 1$. The derivatives of $J$ in (A20) are to be evaluated using the form given in (A4). With (A18) and (A19), one has that the Euler equation for $x_I^{(n)}$ is

$$\frac{\partial L}{\partial x_I^{(n)}} - \frac{d}{d\lambda} \frac{\partial L}{\partial \dot{x}_I^{(n)}} = g_I^{(n)} - p_I^{(n-1)} \Theta_{n-1} - \frac{d}{d\lambda} p_I^{(n)} = 0. \tag{A21}$$

The $p_I^{(n)}$ can be eliminated by differentiating $n$ times and alternately adding and subtracting:

$$\sum_n \left( -\frac{d}{d\lambda} \right)^n \left( g_I^{(n)} - p_I^{(n-1)} \Theta_{n-1} - \frac{d}{d\lambda} p_I^{(n)} \right) = 0 = \sum_n \left( -\frac{d}{d\lambda} \right)^n g_I^{(n)}. \tag{A22}$$

Putting in from (A20)

$$\sum_n \left( -\frac{d}{d\lambda} \right)^n \int d\kappa \left\{ \frac{\partial J}{\partial x_I} \delta_{n,0} + \frac{\partial J}{\partial u_I} \delta_{n,1} + \frac{\kappa^n}{n!} \frac{\partial J}{\partial x_I^+} + \frac{\kappa^{n-1}}{(n-1)!} \frac{\partial J}{\partial u_I^+} \Theta_{n-1} \right\} = 0. \tag{A23}$$

Offsetting by 1 the sum over the last term in braces, $n \to n' = n-1$, (A23) is

$$\int d\kappa \left( \frac{\partial J}{\partial x_I} - \frac{d}{d\lambda} \frac{\partial J}{\partial u_I} \right) + \left[ \sum_n \frac{\kappa^n}{n!} \left( -\frac{d}{d\lambda} \right)^n \right] \int d\kappa \left( \frac{\partial J}{\partial x_I^+} - \frac{d}{d\lambda} \frac{\partial J}{\partial u_I^+} \right) = 0. \tag{A24}$$

Recognizing the expression in square braces as the shift operator

$$\forall h: \quad \sum_n \frac{\kappa^n}{n!} \left( -\frac{d}{d\lambda} \right)^n h(\lambda) = h(\lambda - \kappa), \tag{A25}$$

the second term in (A24) is

$$\left[ \sum_n \frac{\kappa^n}{n!} \left( -\frac{d}{d\lambda} \right)^n \right] \int d\kappa \left( \frac{\partial}{\partial x_I^+} - \frac{d}{d\lambda} \frac{\partial}{\partial u_I^+} \right) J\left( \kappa, x_1, x_2, ..., u_1, u_2, ..., x_1^+, x_2^+, ..., u_1^+, u_2^+, ... \right)$$

$$= \int d\kappa \left( \frac{\partial}{\partial x_I} - \frac{d}{d\lambda} \frac{\partial}{\partial u_I} \right) J\left( \kappa, x_1(\lambda - \kappa), x_2(\lambda - \kappa), ..., u_1(\lambda - \kappa), u_2(\lambda - \kappa), ..., x_1, x_2, ..., u_1, u_2, ... \right)$$

$$= \int d\kappa \left( \frac{\partial}{\partial x_I} - \frac{d}{d\lambda} \frac{\partial}{\partial u_I} \right) J\left( -\kappa, x_1^+, x_2^+, ..., u_1^+, u_2^+, ..., x_1, x_2, ..., u_1, u_2, ... \right) \tag{A26}$$

$$= \int d\lambda' \left( \frac{\partial}{\partial x_I} - \frac{d}{d\lambda} \frac{\partial}{\partial u_I} \right) J\left( \lambda - \lambda', x_1', x_2', ..., u_1', u_2', ..., x_1, x_2, ..., u_1, u_2, ... \right)$$

$$= \int d\lambda' \left( \frac{\partial}{\partial x_I} - \frac{d}{d\lambda} \frac{\partial}{\partial u_I} \right) \tilde{S}(\lambda \leftrightarrow \lambda') J$$

(The third step comes from negating the variable of integration, $\kappa$.) Using this for the second term in (A24) and using that $d\kappa = d\lambda'$ in the first term in (A24), one obtains (A16). It is concluded that the system (A8) is equivalent to the two-time density (A4).

Having now a one-time first order differential form for the action, one can employ the standard results for energy and angular momentum.

**A.4 Restriction to time-monotonic world-lines**

So far the trajectory $x$ has been treated as a general 4-vector without restriction on the $0^{th}$ element. For the remainder of the appendix however, in order to employ established results, it will be necessary to assume that there are *no* time reversals, and therefore that the $0^{th}$ component is a monotonic function of the ordinal parameter $\lambda$. With the $0^{th}$ component no longer a degree of freedom, in place of (A3) and (A4) we now consider the action

$$I = \int dt \int dt'\, K(t' - t, \mathbf{x}_1, \mathbf{x}_2, ..., \mathbf{v}_1, \mathbf{v}_2, ..., \mathbf{x}'_1, \mathbf{x}'_2, ..., \mathbf{v}'_1, \mathbf{v}'_2, ...) = \int dt\, L_K \tag{A27}$$

where

$$L_K \equiv \int d\kappa\, K(\kappa, \mathbf{x}_1, \mathbf{x}_2, ..., \mathbf{v}_1, \mathbf{v}_2, ..., \mathbf{x}_1^+, \mathbf{x}_2^+, ..., \mathbf{v}_1^+, \mathbf{v}_2^+, ...). \tag{A28}$$

Clearly these are not so general as (A3) and (A4) because the time coordinates $t', t$, which could have appeared in an arbitrary way in the latter, must now appear only as a difference. The direct-action kernel (A6) can be written in this form:

$$K = \sum_{j,k} \sum_{\sigma = \pm 1} e_j e_k \left( \mathbf{v}_j \cdot \mathbf{v}_k^+ - 1 \right) \delta\left( \kappa^2 - \left( \mathbf{x}_k^+ - \mathbf{x}_j \right)^2 + \sigma \Delta^2 \right). \tag{A29}$$

The Euler equations corresponding to (A16) are

$$\int dt' \left( \frac{\partial}{\partial \mathbf{x}_I} - \frac{d}{dt} \frac{\partial}{\partial \mathbf{v}_I} \right) \left( 1 + \hat{S}(t \leftrightarrow t') \right) K = 0 \quad \forall I. \tag{A30}$$

The previous analysis demonstrating that (A8) is equivalent to (A4) can be repeated with the action specified by (A27) and (A28) as the starting point. One easily finds that the equivalent one-time system is

$$L = L_{\bar{K}}\left( \mathbf{x}_1^{(0)}, \mathbf{x}_2^{(0)}, ..., \mathbf{x}_1^{(1)}, \mathbf{x}_2^{(1)}, ..., \mathbf{x}_1^{(2)}, \mathbf{x}_2^{(2)}, ..., ..., \right) + L_{aux}\begin{pmatrix} \mathbf{x}_1^{(0)}, \mathbf{x}_2^{(0)}, ..., \mathbf{x}_1^{(1)}, \mathbf{x}_2^{(1)}, ..., \mathbf{x}_1^{(2)}, \mathbf{x}_2^{(2)}, ..., ..., \\ \dot{\mathbf{x}}_1^{(0)}, \dot{\mathbf{x}}_2^{(0)}, ..., \dot{\mathbf{x}}_1^{(1)}, \dot{\mathbf{x}}_2^{(1)}, ..., \dot{\mathbf{x}}_1^{(2)}, \dot{\mathbf{x}}_2^{(2)}, ..., ..., \\ \mathbf{p}_1^{(0)}, \mathbf{p}_2^{(0)}, ..., \mathbf{p}_1^{(1)}, \mathbf{p}_2^{(1)}, ..., \mathbf{p}_1^{(2)}, \mathbf{p}_2^{(2)}, ..., ... \end{pmatrix}, \tag{A31}$$

where

$$L_{aux}(t) = \sum_{j,m} \mathbf{p}_j^{(m)} \cdot \left( \dot{\mathbf{x}}_j^{(m)} - \mathbf{x}_j^{(m+1)} \right). \tag{A32}$$

In place of Eq. (A21) one then has

$$\frac{\partial L}{\partial \mathbf{x}_I^{(n)}} - \frac{d}{dt} \frac{\partial L}{\partial \mathbf{v}_I^{(n)}} = \mathbf{g}_I^{(n)} - \mathbf{p}_I^{(n-1)} \Theta_{n-1} - \frac{d}{dt} \mathbf{p}_I^{(n)} = 0 \tag{A33}$$

where

$$\mathbf{g}_I^{(n)} \equiv \int d\kappa \left\{ \frac{\partial K}{\partial \mathbf{x}_I} \delta_{n,0} + \frac{\partial K}{\partial \mathbf{v}_I} \delta_{n,1} + \frac{\kappa^n}{n!} \frac{\partial K}{\partial \mathbf{x}_I^+} + \frac{\kappa^{n-1}}{(n-1)!} \frac{\partial K}{\partial \mathbf{v}_I^+} \Theta_{n-1} \right\}. \tag{A34}$$

## A.5 Energy of world-lines extremizing a two-time action

Treating the $\mathbf{x}_I^{(n)}$ as independent variables, the Hamiltonian for the system is

$$H = \sum_{l,n} \mathbf{v}_I^{(n)} \cdot \frac{\partial L}{\partial \mathbf{v}_I^{(n)}} - L = \sum_{l,n} \mathbf{v}_I^{(n)} \cdot \mathbf{p}_I^{(n)} - L \tag{A35}$$

where $L$ is the total density. The conjugate momenta $\mathbf{p}_I^{(n)}$ can be found by solving (A33) iteratively as follows. From $n = 0$ onwards (A33) can be written

$$\left(1 + \hat{E}\frac{d}{dt}\right)\mathbf{p}_I^{(n)} = \mathbf{g}_I^{(n+1)} \tag{A36}$$

where $\hat{E}$ increments the superscript: $\hat{E}\mathbf{q}^{(n)} = \mathbf{q}^{(n+1)}, \quad \forall \mathbf{q}^{(n)}$. Assuming convergence, (A36) can be inverted as

$$\mathbf{p}_I^{(n)} = \left(1 - \hat{E}\frac{d}{dt} + \left(\hat{E}\frac{d}{dt}\right)^2 - \left(\hat{E}\frac{d}{dt}\right)^3 + \ldots\right)\mathbf{g}_I^{(n+1)} = \sum_m \left(-\frac{d}{dt}\right)^m \mathbf{g}_I^{(m+n+1)}. \tag{A37}$$

Putting this in (A35) gives

$$\begin{aligned} H &= \sum_{l,m,n} \mathbf{v}_I^{(n)} \cdot \left(-\frac{d}{dt}\right)^m \mathbf{g}_I^{(m+n+1)} - L \\ &= \sum_{l,n,m} \mathbf{v}_I^{(n)} \cdot \left(-\frac{d}{dt}\right)^m \int d\tau \left\{ \frac{\partial K}{\partial \mathbf{x}_I} \delta_{m+n+1,0} + \frac{\partial K}{\partial \mathbf{v}_I} \delta_{m+n+1,1} + \frac{\tau^{m+n+1}}{(m+n+1)!} \frac{\partial K}{\partial \mathbf{x}_I^+} + \frac{\tau^{m+n}}{(m+n)!} \frac{\partial K}{\partial \mathbf{v}_I^+} \Theta_{m+n} \right\} - L \end{aligned} \tag{A38}$$

The first term in braces is zero. The second term is

$$\sum_{l,n,m} \mathbf{v}_I^{(n)} \cdot \left(-\frac{d}{dt}\right)^m \int d\kappa \frac{\partial K}{\partial \mathbf{v}_I} \delta_{m+n+1,1} = \sum_{l,n,m} \mathbf{v}_I^{(n)} \cdot \left(-\frac{d}{dt}\right)^m \int d\kappa \frac{\partial K}{\partial \mathbf{v}_I} \delta_{m,0}\delta_{n,0} = \sum_I \mathbf{v}_I \cdot \int d\kappa \frac{\partial K}{\partial \mathbf{v}_I}. \tag{A39}$$

The total density $L$ appearing in (A38) is the sum (A31). However, the auxiliary density is zero once the Euler equations (A17) are invoked. Therefore $L$ can be replaced with $L_{\bar{K}}$, which is then equal to $L_K$. With this and (A39), (A38) can be decomposed as

$$H = H_0 + H_1 \tag{A40}$$

where

$$H_0 = \sum_I \mathbf{v}_I \cdot \int d\kappa \frac{\partial K}{\partial \mathbf{v}_I} - L_K \tag{A41}$$

and

$$H_1 = \sum_{l,n,m} \mathbf{v}_l^{(n)} \cdot \left(-\frac{d}{dt}\right)^m \int d\kappa \left\{ \frac{\kappa^{m+n}}{(m+n)!} \frac{\partial K}{\partial \mathbf{v}_l^+} + \frac{\kappa^{m+n+1}}{(m+n+1)!} \frac{\partial K}{\partial \mathbf{x}_l^+} \right\}. \tag{A42}$$

(Note that one cannot write (A41) as $H_0 = \sum_l \mathbf{v}_l \cdot \frac{\partial L_K}{\partial \mathbf{v}_l} - L_K$ because the apparent commutivity of the functional derivative with the integration can be destroyed simply by a change of integration variable.) Since it is conserved, the energy and its time-average over $t$ - denoted by braces - are the same. Here we choose to average over a period of the system. In that case the average of any total time derivative will vanish, $\forall g \; \langle dg(t)/dt \rangle = 0$, and therefore

$$\left\langle a(t) \frac{db(t)}{dt} \right\rangle = -\left\langle b(t) \frac{da(t)}{dt} \right\rangle. \tag{A43}$$

With this, (A42) gives

$$\begin{aligned} \langle H_1 \rangle &= \sum_{l,n,m} \left\langle \mathbf{v}_l^{(n)} \cdot \left(-\frac{d}{dt}\right)^m \int d\kappa \left\{ \frac{\kappa^{m+n+1}}{(m+n+1)!} \frac{\partial K}{\partial \mathbf{x}_l^+} + \frac{\kappa^{m+n}}{(m+n)!} \frac{\partial K}{\partial \mathbf{v}_l^+} \right\} \right\rangle \\ &= \sum_{l,n,m} \int d\kappa \left\langle \frac{\kappa^{m+n+1}}{(m+n+1)!} \mathbf{v}_l^{(n+m)} \cdot \left( \frac{\partial K}{\partial \mathbf{x}_l^+} - \frac{d}{d\kappa} \frac{\partial K}{\partial \mathbf{v}_l^+} \right) \right\rangle \end{aligned} \tag{A44}$$

where the second term in braces has been integrated over $\kappa$ by parts, and it is assumed that all quantities vanish on the boundary (the limits of the integration over $\kappa$). Letting $m+n=p$ and replacing the sum over $m$, the above becomes

$$\langle H_1 \rangle = \sum_{l,p} \sum_{n=0}^{p} \int d\kappa \left\langle \frac{\kappa^{p+1}}{(p+1)!} \mathbf{v}_l^{(p)} \cdot \left( \frac{\partial K}{\partial \mathbf{x}_l^+} - \frac{d}{d\kappa} \frac{\partial K}{\partial \mathbf{v}_l^+} \right) \right\rangle. \tag{A45}$$

Noting that the sum over $n$ just gives $p+1$ and using (A25), this is

$$\langle H_1 \rangle = \sum_{l,p} \int d\kappa \, \kappa \left\langle \frac{\kappa^p}{p!} \mathbf{v}_l^{(p)} \cdot \left( \frac{\partial K}{\partial \mathbf{x}_l^+} - \frac{d}{d\kappa} \frac{\partial K}{\partial \mathbf{v}_l^+} \right) \right\rangle = \sum_l \int d\kappa \, \kappa \left\langle \mathbf{v}_l^+ \cdot \left( \frac{\partial K}{\partial \mathbf{x}_l^+} - \frac{d}{d\kappa} \frac{\partial K}{\partial \mathbf{v}_l^+} \right) \right\rangle. \tag{A46}$$

Noting that

$$\frac{dK}{d\kappa} = \sum_l \left\{ \mathbf{v}_l^+ \cdot \frac{\partial K}{\partial \mathbf{x}_l^+} + \dot{\mathbf{v}}_l^+ \cdot \frac{\partial K}{\partial \mathbf{v}_l^+} \right\} + \frac{\partial K}{\partial \kappa} = \sum_l \left\{ \mathbf{v}_l^+ \cdot \left( \frac{\partial K}{\partial \mathbf{x}_l^+} - \frac{d}{d\kappa} \frac{\partial K}{\partial \mathbf{v}_l^+} \right) + \frac{d}{d\kappa} \left( \mathbf{v}_l^+ \cdot \frac{\partial K}{\partial \mathbf{v}_l^+} \right) \right\} + \frac{\partial K}{\partial \kappa}, \tag{A47}$$

(A46) is

$$\langle H_1 \rangle = \int d\kappa \, \kappa \left\langle \frac{dK}{d\kappa} - \frac{\partial K}{\partial \kappa} - \sum_l \frac{d}{d\kappa} \left( \mathbf{v}_l^+ \cdot \frac{\partial K}{\partial \mathbf{v}_l^+} \right) \right\rangle. \tag{A48}$$

Integrating by parts and again assuming that quantities vanish on the boundary:

$$\langle H_1 \rangle = \int d\kappa \left\langle \sum_l \left( \mathbf{v}_l^+ \cdot \frac{\partial K}{\partial \mathbf{v}_l^+} \right) - K - \kappa \frac{\partial K}{\partial \kappa} \right\rangle = \left\langle \sum_l \int d\kappa \, \mathbf{v}_l^+ \cdot \frac{\partial K}{\partial \mathbf{v}_l^+} - L_K - \int d\kappa \, \kappa \frac{\partial K}{\partial \kappa} \right\rangle. \tag{A49}$$

Combining (A41) and (A49), and, for notational simplicity, extending the time average to include the terms in $H_0$, one finally obtains

$$H = \sum_I \int d\kappa \left\langle \mathbf{v}_I^+ \cdot \frac{\partial K}{\partial \mathbf{v}_I^+} \right\rangle + \int d\kappa \sum_I \left\langle \mathbf{v}_I \cdot \frac{\partial K}{\partial \mathbf{v}_I} \right\rangle - 2\langle L_K \rangle - \int d\kappa\, \kappa \left\langle \frac{\partial K}{\partial \kappa} \right\rangle \qquad (A50)$$

where the angle brackets denote an average over a period of *t*.

The plausibility of (A50) can be demonstrated by showing that the expression Eq. (A50) agrees with the traditional result when the action is one-time. If one sets

$$K = \delta(\kappa) L_{trad}(\mathbf{x}, \mathbf{v}) \qquad (A51)$$

then (A27) with (A28) gives the traditional form

$$I = \int dt\, L_{trad}(\mathbf{x}, \mathbf{v}). \qquad (A52)$$

Putting (A51) in (A50) and noting that $\partial L_{trad}/\partial \mathbf{v}^+ = 0$, one has

$$H = \left\langle \mathbf{v} \cdot \frac{\partial L_{trad}}{\partial \mathbf{v}} \right\rangle - 2\langle L_{trad} \rangle - \int d\kappa\, \kappa \left\langle \frac{\partial \delta(\kappa)}{\partial \kappa} L_{trad}(\mathbf{x}, \mathbf{v}) \right\rangle. \qquad (A53)$$

Noting now that in this case the partial derivative with respect to *κ* is the same as the total derivative with respect to *κ*, the last term can be integrated by parts, after which one simply has

$$H = \left\langle \mathbf{v} \cdot \frac{\partial L_{trad}}{\partial \mathbf{v}} \right\rangle - \langle L_{trad} \rangle \qquad (A54)$$

as required.

### A.6 Energy of charges obeying direct-action EM

Computed above is the energy associated with a general two-time action wherein the explicit dependence on time is only on the time difference, i.e.

$$I = \int dt \int dt'\, K(t'-t, \mathbf{x}_1(t), \mathbf{x}_2(t),\ldots, \mathbf{v}_1(t), \mathbf{v}_2(t),\ldots, \mathbf{x}_1(t'), \mathbf{x}_2(t'),\ldots, \mathbf{v}_1(t'), \mathbf{v}_2(t'),\ldots). \qquad (A55)$$

Provided the particle world lines are time-monotonic, it is shown that energy is given by (A50):

$$H = \int dt' \left\langle \sum_I \left\{ \mathbf{v}_I' \cdot \frac{\partial K}{\partial \mathbf{v}_I'} + \mathbf{v}_I \cdot \frac{\partial K}{\partial \mathbf{v}_I} \right\} - 2K - (t'-t)\frac{\partial K}{\partial t'} \right\rangle \qquad (A56)$$

where

$$\mathbf{x}_j \equiv \mathbf{x}_j(t), \quad \mathbf{v}_j \equiv \mathbf{v}_j(t), \quad \mathbf{x}_j' \equiv \mathbf{x}_j(t'), \quad \mathbf{v}_j' \equiv \mathbf{v}_j(t') \qquad (A57)$$

where the angle brackets signify a time-average (over *t*) over a period of the motion. In the particular case of the direct-action EM considered in this document, the *K* in (A55) implied by (3) is

$$K = \sum_{\sigma_s = \pm 1} \sum_{j,k} e_j e_k (\mathbf{v}_j \cdot \mathbf{v}_k' - 1)\delta\!\left((t'-t)^2 - (\mathbf{x}_k' - \mathbf{x}_j)^2 + \sigma_s \Delta^2\right). \qquad (A58)$$

Inserting (A58) into (A56), the energy a system of time-monotonic, self-interacting, mechanically massless particles interacting electromagnetically is found to be

$$H = \sum_{\sigma_s = \pm 1} \sum_{j,k} e_j e_k \int dt' \left\langle \left( 2 - (t'-t)(\mathbf{v}_j \cdot \mathbf{v}'_k - 1) \frac{\partial}{\partial t'} \right) \delta\left( (t'-t)^2 - (\mathbf{x}'_k - \mathbf{x}_j)^2 + \sigma_s \Delta^2 \right) \right\rangle. \quad \text{(A59)}$$

Noting that

$$\frac{d}{dt'} \delta\left( (t'-t)^2 - (\mathbf{v}'_k - \mathbf{x}_j)^2 + \sigma_s \Delta^2 \right) = 2\left( t' - t - \mathbf{v}'_k \cdot (\mathbf{x}'_k - \mathbf{x}_j) \right) \delta'\left( (t'-t)^2 - (\mathbf{x}'_k - \mathbf{x}_j)^2 + \sigma_s \Delta^2 \right). \quad \text{(A60)}$$

then

$$\frac{\partial}{\partial t'} \delta\left( (t'-t)^2 - (\mathbf{x}^+_k - \mathbf{x}_j)^2 + \sigma_s \Delta^2 \right) = 2(t'-t) \delta'\left( (t'-t)^2 - (\mathbf{x}'_k - \mathbf{x}_j)^2 + \sigma_s \Delta^2 \right)$$
$$= \frac{t'-t}{t' - t - \mathbf{v}'_k \cdot (\mathbf{x}'_k - \mathbf{x}_j)} \frac{d}{d\kappa} \delta\left( (t'-t)^2 - (\mathbf{x}'_k - \mathbf{x}_j)^2 + \sigma_s \Delta^2 \right). \quad \text{(A61)}$$

Putting this in (A59) and integrating the second term by parts gives

$$H = \sum_{\sigma_s = \pm 1} \sum_{j,k} e_j e_k \int dt' \left\langle \left( 2 + \frac{d}{dt'} \left[ \frac{(t'-t)^2 (\mathbf{v}_j \cdot \mathbf{v}'_k - 1)}{t' - t - \mathbf{v}'_k \cdot (\mathbf{x}'_k - \mathbf{x}_j)} \right] \right) \delta\left( (t'-t)^2 - (\mathbf{x}'_k - \mathbf{x}_j)^2 + \sigma_s \Delta^2 \right) \right\rangle. \quad \text{(A62)}$$

Noting that the result of a total differential is insensitive to the choice of algebraic representation of its operand, the 4-vector notation can be restored to give

$$H = \sum_{\sigma_s = \pm 1} \sum_{j,k} e_j e_k \int dt' \left\langle \left( 2 - \frac{d}{dt'} \left[ \frac{(t'-t)^2 u'_k \circ u_j}{u'_k \circ s_{k,j}} \right] \right) \delta\left( s^2_{k,j} + \sigma_s \Delta^2 \right) \right\rangle. \quad \text{(A63)}$$

Carrying out the integration over $t'$ and performing the differentiation, this is

$$H = \frac{1}{2} \sum_{\sigma_s = \pm 1} \sum_{j,k} \sum_{roots} \left\langle \frac{e_j e_k}{|u'_k \circ s_{k,j}|^3} \begin{pmatrix} 2(u'_k \circ s_{k,j})^2 - 2(t'-t)(u'_k \circ s_{k,j})(u'_k \circ u_j) \\ + (t'-t)^2 \left( (u^+_k \circ u_j)(u'^2_k + a'_k \circ s_{k,j}) - (u'_k \circ s_{k,j})(a'_k \circ u_j) \right) \end{pmatrix} \right\rangle. \quad \text{(A64)}$$

where the roots are those values of $t'$ satisfying (17). Eq. (A64) is a generally valid expression for the total energy of a system of time-monotonic mechanically massless particles interacting according to direct-action EM. It is evaluated in the particular case of dual circular motion in the following section.

### A.7 Angular momentum of world-lines extremizing a two-time action

Treating the $\mathbf{x}^{(n)}_l$ as independent variables, the angular momentum $\mathbf{L}$ of the density (A31) is

$$\mathbf{L} = \sum_{l,n} \mathbf{x}^{(n)}_l \times \frac{\partial L}{\partial \dot{\mathbf{x}}^{(n)}_l} = \sum_{l,n} \mathbf{x}^{(n)}_l \times \mathbf{p}^{(n)}_l. \quad \text{(A65)}$$

Using the result (A37) for the conjugate momentum and inserting (A34) this is

$$\mathbf{L} = \sum_{l,m,n} \mathbf{x}^{(n)}_l \times \left( -\frac{d}{dt} \right)^m \int d\kappa \left\{ \frac{\partial K}{\partial \mathbf{x}_l} \delta_{m+n+1,0} + \frac{\partial K}{\partial \mathbf{v}_l} \delta_{m+n+1,1} + \frac{\kappa^{m+n+1}}{(m+n+1)!} \frac{\partial K}{\partial \mathbf{x}^+_l} + \frac{\kappa^{m+n}}{(m+n)!} \frac{\partial K}{\partial \mathbf{v}^+_l} \Theta_{m+n} \right\}. \quad \text{(A66)}$$

We now proceed much as for the energy. The first term in braces is zero. The second term is

$$\sum_{l,m,n} \mathbf{x}_l^{(n)} \times \left(-\frac{d}{dt}\right)^m \int d\kappa \, \frac{\partial K}{\partial \mathbf{v}_l} \delta_{m+n+1,1} = \sum_{l,m,n} \mathbf{x}_l^{(n)} \times \left(-\frac{d}{dt}\right)^m \int d\kappa \, \frac{\partial K}{\partial \mathbf{v}_l} \delta_{m,0} \delta_{n,0} = \sum_l \mathbf{x}_l \times \int d\kappa \, \frac{\partial K}{\partial \mathbf{v}_l}. \tag{A67}$$

Using (A67) we can write for (A66)

$$\mathbf{L} = \mathbf{L}_0 + \mathbf{L}_1 \tag{A68}$$

where

$$\mathbf{L}_0 = \sum_l \mathbf{x}_l \times \int d\kappa \, \frac{\partial K}{\partial \mathbf{v}_l} \tag{A69}$$

and

$$\mathbf{L}_1 = \sum_{l,m,n} \mathbf{x}_l^{(n)} \times \left(-\frac{d}{dt}\right)^m \int d\kappa \left\{ \frac{\kappa^{m+n+1}}{(m+n+1)!} \frac{\partial K}{\partial \mathbf{x}_l^+} + \frac{\kappa^{m+n}}{(m+n)!} \frac{\partial K}{\partial \mathbf{v}_l^+} \right\}. \tag{A70}$$

Since it is conserved, the angular momentum and therefore its time-average over $t$ - denoted in the following by angle-brackets - are the same. Repeating steps leading from (A44) to (A46) then gives

$$\langle \mathbf{L}_1 \rangle = \sum_l \int d\kappa \, \kappa \left\langle \mathbf{x}_l^+ \times \left( \frac{\partial K}{\partial \mathbf{x}_l^+} - \frac{d}{d\kappa} \frac{\partial K}{\partial \mathbf{v}_l^+} \right) \right\rangle. \tag{A71}$$

Integrating the second term by parts:

$$\langle \mathbf{L}_1 \rangle = \sum_l \int d\kappa \left\langle \mathbf{x}_l^+ \times \frac{\partial K}{\partial \mathbf{v}_l^+} + \kappa \left( \mathbf{x}_l^+ \times \frac{\partial K}{\partial \mathbf{x}_l^+} + \mathbf{v}_l^+ \times \frac{\partial K}{\partial \mathbf{v}_l^+} \right) \right\rangle. \tag{A72}$$

Combining this with (A69) one has that the angular momentum of a system with action (A27) with (A28) is

$$\mathbf{L} = \sum_l \int d\kappa \left\langle \mathbf{x}_l \times \frac{\partial K}{\partial \mathbf{v}_l} + \mathbf{x}_l^+ \times \frac{\partial K}{\partial \mathbf{v}_l^+} + \kappa \left( \mathbf{x}_l^+ \times \frac{\partial K}{\partial \mathbf{x}_l^+} + \mathbf{v}_l^+ \times \frac{\partial K}{\partial \mathbf{v}_l^+} \right) \right\rangle. \tag{A73}$$

### A.8 Angular momentum of charges obeying direct-action EM

The general result for the angular momentum vector of a closed system whose action is of the form (A55) is

$$\mathbf{L} = \sum_l \int dt' \left\langle \mathbf{x}_l \times \frac{\partial K}{\partial \mathbf{v}_l} + \mathbf{x}_l' \times \frac{\partial K}{\partial \mathbf{v}_l'} + (t' - t) \left( \mathbf{x}_l' \times \frac{\partial K}{\partial \mathbf{x}_l'} + \mathbf{v}_l' \times \frac{\partial K}{\partial \mathbf{v}_l'} \right) \right\rangle. \tag{A74}$$

Using the $K$ introduced in (A58) the angular momentum of a system of time-monotonic, self-interacting - but mechanically massless - particles interacting electromagnetically is

$$\mathbf{L} = \sum_{j,k} \sum_{\sigma_s = \pm 1} e_j e_k \int dt' \left\langle \begin{array}{c} \left( \mathbf{x}_j \times \mathbf{v}_k' + \left( \mathbf{x}_k' + (t'-t)\mathbf{v}_k' \right) \times \mathbf{v}_j + (t'-t)\left( \mathbf{v}_k'.\mathbf{v}_j - 1 \right) \mathbf{x}_k' \times \frac{\partial}{\partial \mathbf{x}_k'} \right) \\ \times \delta\left( (t'-t)^2 - \left( \mathbf{x}_k' - \mathbf{x}_j \right)^2 + \sigma_s \Delta^2 \right) \end{array} \right\rangle. \tag{A75}$$

Using that

$$\frac{\partial}{\partial \mathbf{x}'_k}\delta\!\left((t'-t)^2-\left(\mathbf{x}'_k-\mathbf{x}_j\right)^2+\sigma_s\Delta^2\right)=-2\left(\mathbf{x}'_k-\mathbf{x}_j\right)\delta'\!\left((t'-t)^2-\left(\mathbf{x}'_k-\mathbf{x}_j\right)^2+\sigma_s\Delta^2\right)$$

$$=-\frac{\mathbf{x}'_k-\mathbf{x}_j}{t'-t-\mathbf{v}'_k\cdot\left(\mathbf{x}'_k-\mathbf{x}_j\right)}\frac{d}{dt'}\delta\!\left((t'-t)^2-\left(\mathbf{x}'_k-\mathbf{x}_j\right)^2+\sigma_s\Delta^2\right), \qquad \text{(A76)}$$

and assuming all relevant quantities vanish on the boundary, (A75) can be written

$$\mathbf{L}=\sum_{j,k}\sum_{\sigma_s=\pm 1}e_j e_k\int dt'\left\langle\begin{pmatrix}\mathbf{x}_j\times\mathbf{v}'_k+\left(\mathbf{x}'_k+(t'-t)\mathbf{v}'_k\right)\times\mathbf{v}_j-\dfrac{d}{dt'}\!\left(\dfrac{(t'-t)(\mathbf{v}'_k\cdot\mathbf{v}_j-1)\mathbf{x}'_k\times\mathbf{x}_j}{t'-t-\mathbf{v}'_k\cdot\left(\mathbf{x}'_k-\mathbf{x}_j\right)}\right)\\ \times\delta\!\left((t'-t)^2-\left(\mathbf{x}'_k-\mathbf{x}_j\right)^2+\sigma_s\Delta^2\right)\end{pmatrix}\right\rangle$$

$$=\sum_{j,k}\sum_{\sigma_s=\pm 1}\sum_{roots}\frac{e_j e_k}{2}\left\langle\frac{1}{|u'_k\circ s_{k,j}|}\left(\mathbf{x}_j\times\mathbf{v}'_k+\left(\mathbf{x}'_k+(t'-t)\mathbf{v}'_k\right)\times\mathbf{v}_j+\frac{d}{dt'}\!\left(\frac{(t'-t)(u'_k\circ u_j)\mathbf{x}'_k\times\mathbf{x}_j}{u'_k\circ s_{k,j}}\right)\right)\right\rangle. \qquad \text{(A77)}$$

$$=\sum_{j,k}\sum_{\sigma_s=\pm 1}\sum_{roots}\frac{e_j e_k}{2}\left\langle\frac{1}{|u'_k\circ s_{k,j}|^3}\begin{pmatrix}\left(u'_k\circ s_{k,j}\right)^2\left(\mathbf{x}_j\times\mathbf{v}'_k+\left(\mathbf{x}'_k+(t'-t)\mathbf{v}'_k\right)\times\mathbf{v}_j\right)\\ +(t'-t)\left(u'_k\circ s_{k,j}\right)\left(u'_k\circ u_j\right)\mathbf{v}'_k\times\mathbf{x}_j\\ +\begin{pmatrix}\left(u'_k\circ s_{k,j}\right)\left(u'_k\circ u_j+(t'-t)a'_k\circ u_j\right)\\ -(t'-t)\left(u'^2_k+a'_k\circ s_{k,j}\right)\left(u'_k\circ u_j\right)\end{pmatrix}\mathbf{x}'_k\times\mathbf{x}_j\end{pmatrix}\right\rangle$$

## References

[1] F. Rohrlich, Phys. Rev. D 60 (1999) 084017-084017-5.

[2] T. Erber, Fortschritte der Physik 9 (1961) 343.

[3] A. D. Fokker, Zeits. f. Phys. 58 (1929) 386.

[4] K. Schwarzschild, Gottinger Nachrichten 128 (1903) 132.

[5] H. Tetrode, Zeits. f. Phys. 10 (1922) 317.

[6] J. A. Wheeler and R. P. Feynman, Rev. Mod. Phys. 17 (1945) 157.

[7] J. A. Wheeler and R. P. Feynman, Rev. Mod. Phys. 21 (1949) 425.

[8] P. C. W. Davies, The Physics of Time Asymmetry, University of California Press, 1977.

[9] P. C. W. Davies, J. Phys. A 5 (1972) 1722.

[10] D. T. Pegg, Rep. Prog. Phys. 38 (1975) 1339.

[11] M. Ibison, Fizika A 12 (2004) 55.

[12] F. Hoyle and J. V. Narlikar, Proc. Roy. Soc. Lond. A 277 (1964) 1.

[13] J. E. Hogarth, Proc. Roy. Soc. Lond. A 267 (1962) 365.

[14] F. Hoyle and J. V. Narlikar, Annals of Physics 54 (1969) 207.

[15] P. C. W. Davies, Proc. Cam. Phil. Soc. 68 (1970) 751.

[16] D. T. Pegg, Phys. Lett. A 76 (1980) 109.

[17] R. P. Feynman, Phys. Rev. 76 (1949) 749.

[18] M. Ibison, in: A. Chubykalo et al (Eds.) Has the Last Word Been Said on Classical Electrodynamics?, Rinton Press, 2004.

[19] F. Hoyle and J. V. Narlikar, Annals of Physics 62 (1971) 44.

[20] F. Hoyle and J. V. Narlikar, Action at a Distance in Physics and Cosmology, W. H. Freeman, 1974.

[21] P. C. W. Davies, J. Phys. A 5 (1972) 1024.

[22] J. D. Jackson, Classical Electrodynamics, John Wiley & Sons, Inc., 1998.

[23] L. D. Landau and E. M. Lifshitz, The Classical Theory of Fields, Pergamon Press, 1980.

[24] A. Sommerfeld, Proceedings of the Amsterdam Academy 8 (1904) 346.

[25] P. Cerenkov, I. Tamm, and I. Frank, Sov. Phys. , Usp. 68 (1958) 376.

[26] J. L. Augdin and A. M. Platzeck, Phys. Rev. D 26 (1982) 1923.

[27] F. C. Jones, Phys. Rev. D 6 (1972) 2727.

[28] H. K. Wimmel, Nature Phys. Sci. 236 (1972) 79.

[29] M. Fayngold, Special Relativity and Motions Faster than Light, Wiley-VCH, 2002.

[30] E. Recami and R. Mignani, Rivista del Nuovo Cimento 4 (1974) 209.

[31] E. Recami Tachyons, Monopoles, and Related Topics, North-Holland, 1978.

[32] H. C. Corben, in: E. Recami (Ed.) Tachyons, Monopoles and Related Topics, North-Holland, 1978.

[33] R. Mignani and E. Recami, Il Nuovo Cimento 14A (1973) 169.

[34] O. M. Bilaniuk, V. K. Deshpande, and E. C. G. Sudarshan, Am. J. Phys. 30 (1962) 718.

[35] A. O. Barut, Electrodynamics and Classical Theory of Fields and Particles, Dover, 1980.

**Table 1**

Title: Solutions of the equation $\beta_0 \tan \beta_0 = -1$ with $|\mathbf{v}_0| = \sqrt{\beta_0^2 + 1}$.

| mode index | $\beta_0$ | $|\mathbf{v}_0|$ |
|---|---|---|
| 1 | 2.798 | 2.972 |
| 2 | 6.121 | 6.202 |
| 3 | 9.318 | 9.371 |
| 4 | 12.486 | 12.526 |
| 5 | 15.644 | 15.676 |
| 6 | 18.796 | 18.823 |
| 7 | 21.946 | 21.968 |
| large $n$ | $n\pi$ | $n\pi$ |

**Figure Captions**

**Figure 1a**

Effect of modified light cone condition on points of interaction. The present position of the particle is at the origin. The two red dots show the points of local self-action. The blue dots show the points of distant action for arbitrary motion of a distant trajectory.

**Figure 1b**

The same modified light cone condition except now the distant trajectory intersects the light cone at a tangent so there is just one point of interaction with each segment. In 1+1D tangency to the light cone necessitates light-speed at the point of contact, but in 2+1D and 3+1D the tangency condition can be fulfilled by a distant trajectory having any superluminal speed.

**Figure 2a**

Forces keeping the two charges in circular motion. A positive charge of infinite electromagnetic mass, here denoted by the red disk, lies – simultaneously - on the light cone and the Cerenkov cone of both the points indicated by dark blue disks, which are historical and future locations of the negative charge. The resultant force on the positive charge is singular and directed towards the origin. An identical relationship exists between the present location of the negative charge and the historical and future locations of the positive charge.

**Figure 2b**

Details of the geometry showing the role of the angle $\beta$ for the mode $n = 1$. A charge moves through an angle $2\beta$ during the time it takes for the singular electromagnetic contact to propagate to its opposite signed partner. For the mode $n = 1$ this is 5.596 rad. $= 320.6^{\circ}$.

**Figure 3a**

A rendering of the circular motion in 2+1D for the mode $n = 1$ showing points of singular electromagnetic contact. The lower / upper brown dots, for example, show the location of positive charge when it was / will be simultaneously on the light-cone and Cerenkov cone of the present location of the negative charge.

**Figure 3b**

A rendering of the circular motion in 2+1D for the mode $n = 1$ showing the interaction points on the light cone of the positive charge. The pink dots are non-singular distant self-interactions. The light-blue dots (only one of which is visible in this view) are non-singular distant interactions with the other – negatively-charged – particle. The dark blue dots are the singular distant interactions with the negatively-charged particle, the space-time helical path of which grazes the light cone at these points in fulfillment of the Cerenkov cone condition. The heavy black lines show the locus of the singular interactions.

**Figure 1a**

Effect of modified light cone condition on points of interaction. The present position of the particle is at the origin. The two red dots show the points of local self-action. The blue dots show the points of distant action for arbitrary motion of a distant trajectory.

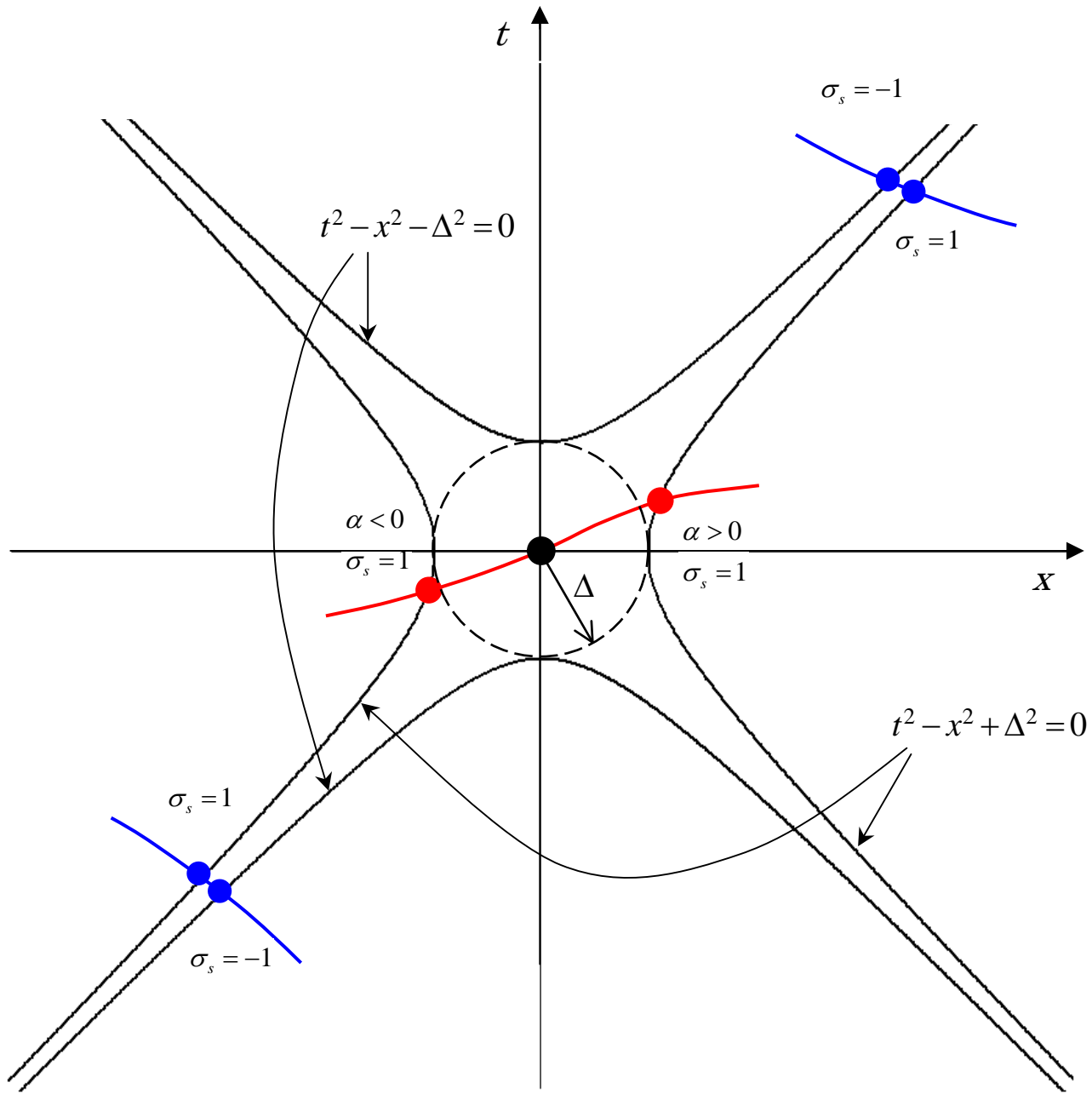

**Figure 1b**
The same modified light cone condition except now the distant trajectory intersects the light cone at a tangent so there is just one point of interaction with each segment. In 1+1D tangency to the light cone necessitates light-speed at the point of contact, but in 2+1D and 3+1D the tangency condition can be fulfilled by a distant trajectory having any superluminal speed.

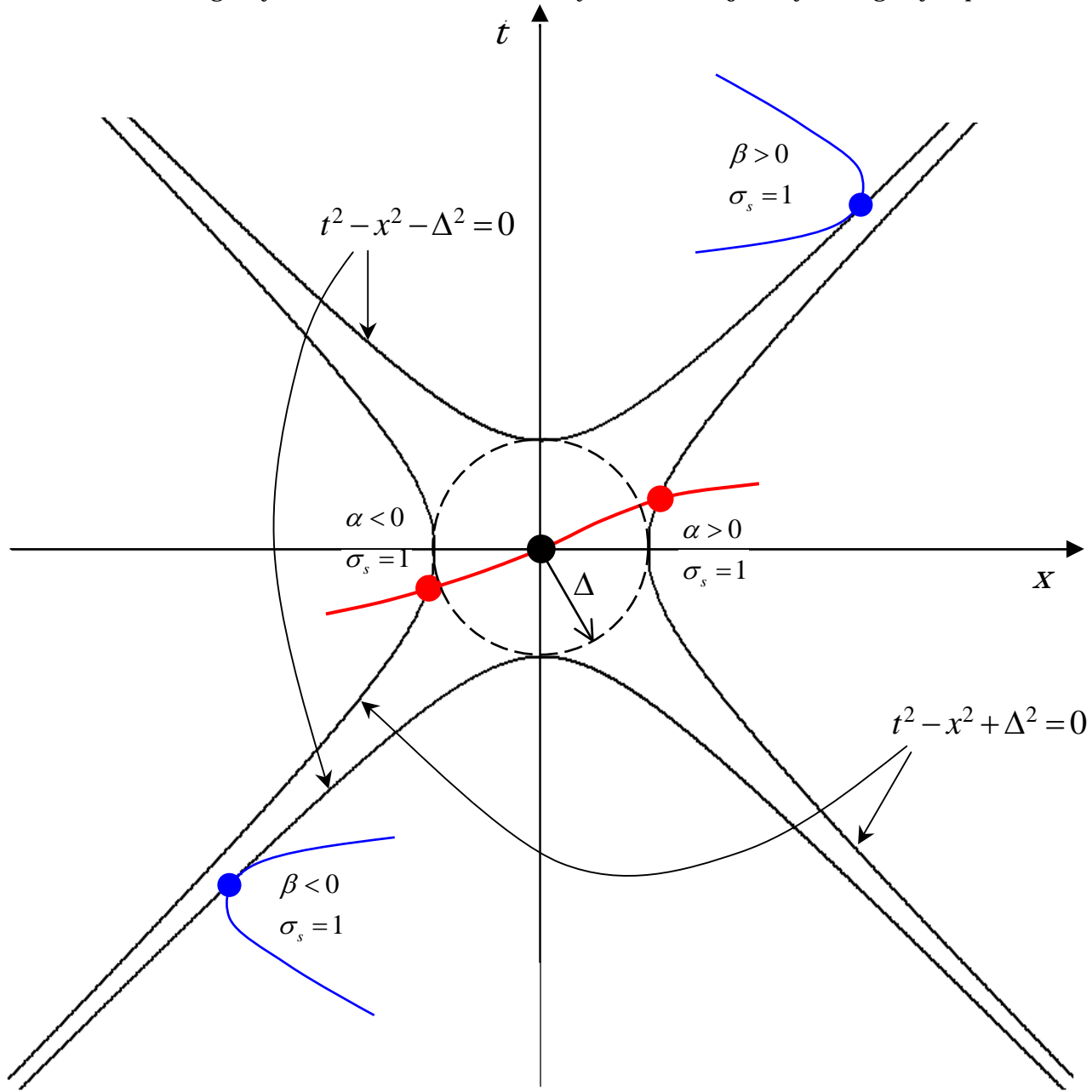

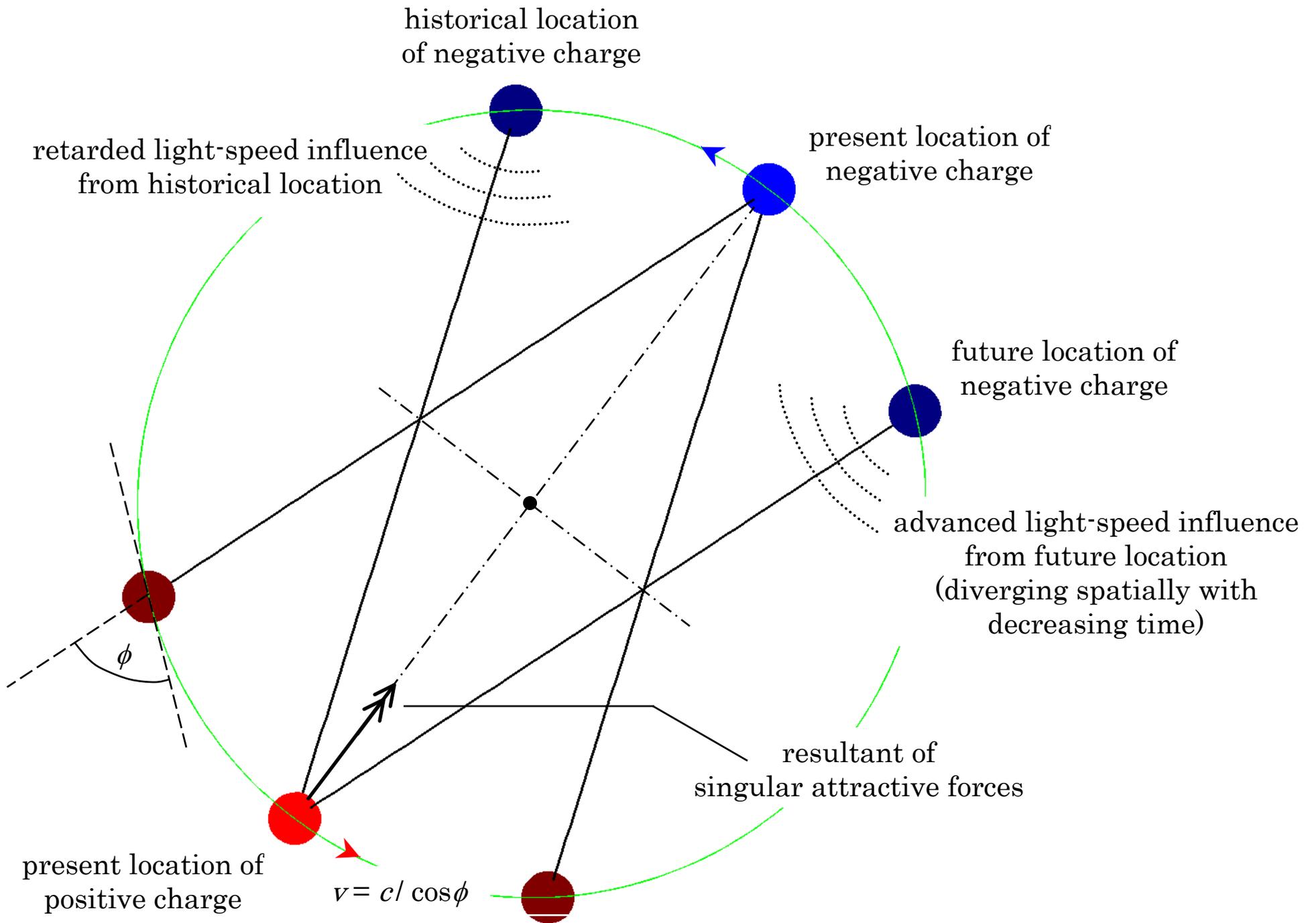

**Figure 2a**

Forces keeping the two charges in circular motion. A positive charge of infinite electromagnetic mass, here denoted by the red disk, lies - simultaneously - on the light cone and the Cerenkov cone of both the points indicated by dark blue disks, which are historical and future locations of the negative charge.

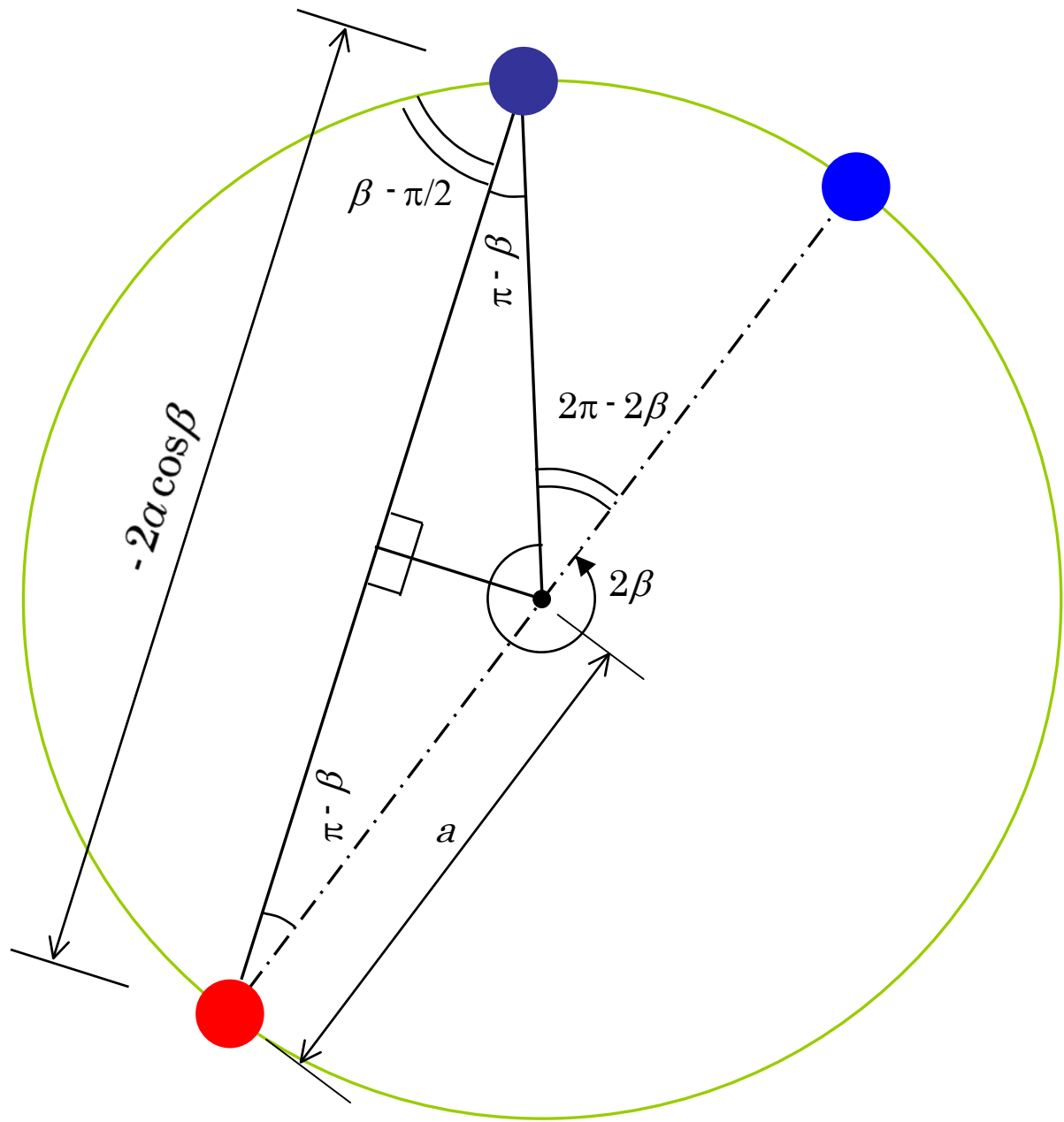

**Figure 2b**

Details of the geometry showing the role of the angle β for the mode n = 1. A charge moves through an angle 2β during the time it takes for the singular electromagnetic contact to propagate to its opposite signed partner.

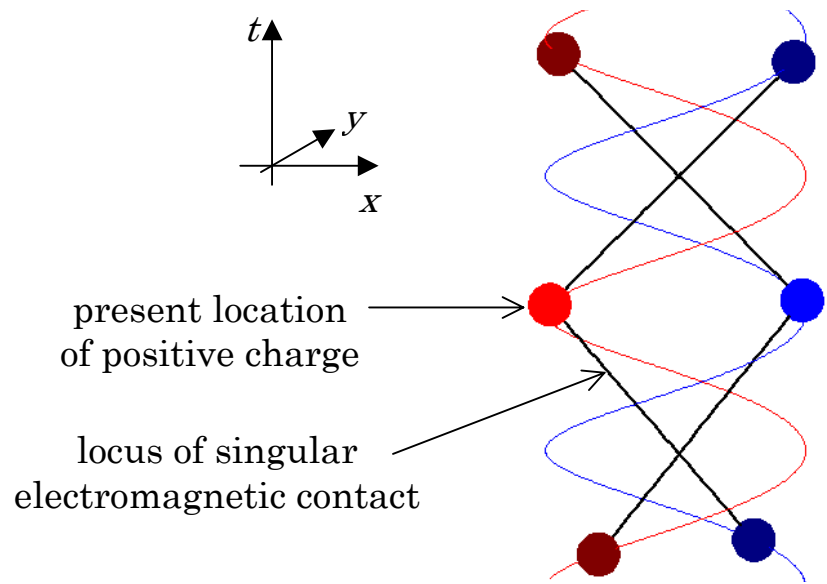

**Figure 3a**

A rendering of the circular motion in 2+1D for the mode n = 1 showing points of singular electromagnetic contact. The lower / upper brown dots, for example, show the location of positive charge when it was / will be simultaneously on the light-cone and Cerenkov cone of the present location of the negative charge.

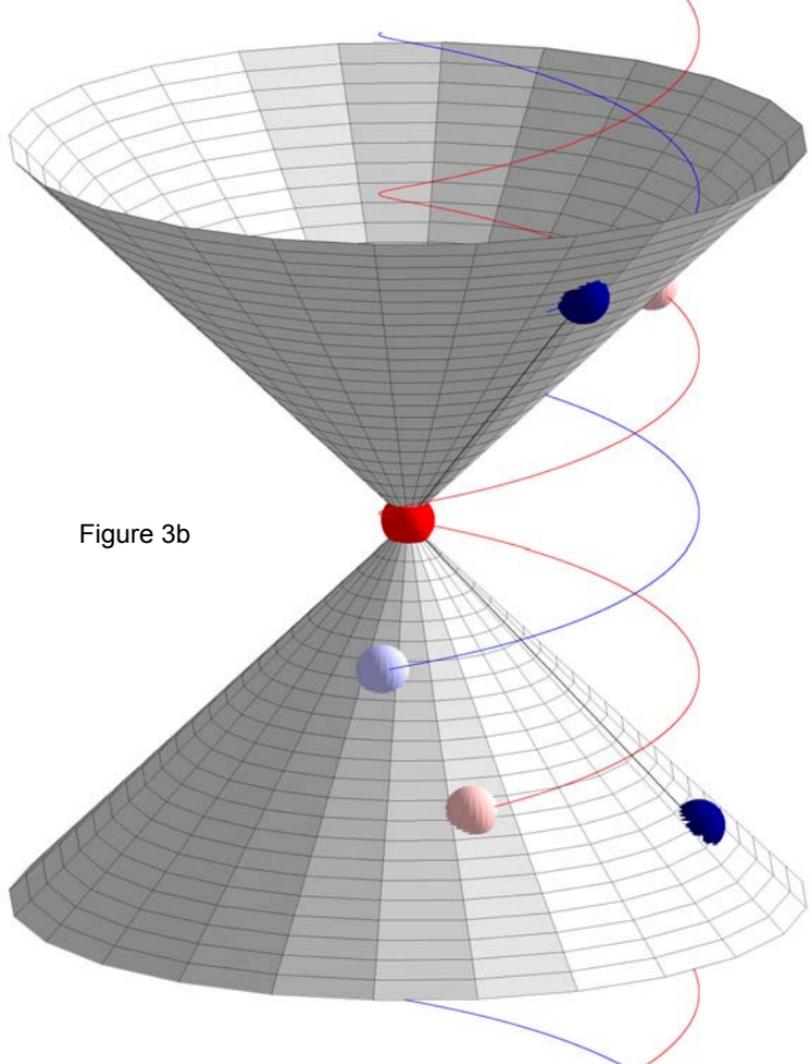

Figure 3b